\documentstyle[prl,preprint,aps,psfig]{revtex}
\begin{document}
\draft

\title{Roton Instability of the Spin Wave Excitation in the Fully Polarized
Quatum Hall State and the Phase Diagram at $\nu = 2$}
\author{K. Park and J. K. Jain}
\address{Department of Physics, 104 Davey Laboratory,
The Pennsylvania State University,
University park,
Pennsylvania 16802}
\date{\today}

\maketitle

\begin{abstract}

We consider the effect of interactions on electrons confined to two 
dimensions at Landau level filling $\nu=2$, with the specific aim to 
determine the range of parameters where the fully polarized state is stable.
We calculate the
charge and the spin density collective modes in random phase approximation
(RPA) including vertex corrections (also known as time dependent Hartree
Fock), and treating the Landau level mixing accurately within the 
subspace of a
single particle hole pair.  
It is found that the spin wave excitation mode
of the fully polarized state 
has a roton minimum which deepens as a result of 
the interaction induced Landau level
mixing, and the energy of the roton 
vanishes at a critical Zeeman energy signaling
an instability of the fully polarized state at still lower Zeeman energies.
The feasibility of the experimental observation of the roton minimum in the
spin wave mode and its softening will be discussed.
The spin and charge density collective modes of the unpolarized state are 
also considered, and a phase diagram for the $\nu =2$ state  
as a function of $r_{S}$ and the Zeeman energy is obtained.

\end{abstract}


\section{Introduction}
The interplay between the electron's spin degree of freedom 
and the inter-electron interaction 
has been of interest in the condensed matter physics, in particular in
the jellium model where the neutralizing background is taken to be
rigid and uniform.  
The relative strength of the interaction is conventionally
measured through the parameter $r_S$ which
is the interparticle distance measured in units of the 
Bohr radius $a_B \equiv \epsilon \hbar^2/m e^2$, $\epsilon$ being the
dielectric constant and $m$ the band mass of electron.
For electrons confined to two dimensions, which will be our 
focus in this paper, we have  
\begin{equation}
r_S\equiv \frac{(\pi \rho)^{-1/2}}{a_B}\;\; ,
\end{equation}
where  $\rho$ is the two-dimensional density of electrons.
The interaction strength is
enhanced relative to the kinetic energy as the system becomes more
dilute, i.e., as $r_S$ increases.
It has been predicted that as $r_S$ is increased,
the electron liquid eventually 
becomes spontaneously spin polarized to gain in the exchange energy, ultimately
going into a Wigner Crystal at $r_S\approx 37$ \cite{Ceperley}.

The two-dimensional electron systems, 
obtained experimentally at the interface of two semiconductors, 
constitute an almost ideal realization of the jellium model 
for several reasons.  Samples with mobility in access of 10 million 
cm$^2$/Vs are available \cite{Pan}, minizing the effect of disorder.
Furthermore, the density of electrons, which controls the 
strength of the interaction relative to the kinetic energy, can be varied
by a factor of 20 \cite{Pan,Pinczuk}. 
We will consider electrons in the presence of
a magnetic field, specifically, at filling factor $\nu=2$, which is a
particularly clean test case for the kind of physics in which 
we are interested.  There are three relevant energy scales here:
$\hbar \omega_C=\hbar e B/m c$ is the cyclotron
energy, $V_C=e^2/\epsilon l_0$ is the typical Coulomb energy,
$l_0=\sqrt{\hbar c/e B}$ being the magnetic length, and $E_Z$ is the
Zeeman splitting energy which is the Zeeman
energy cost necessary for a single spin flip.   
(To an extent, the Zeeman and the cyclotron energies can be varied
independently by application of the magnetic field at an angle; while the
former depends on the total magnetic field, the latter is determined by the
normal component only.) 
The ground state is known in two limits. When $\hbar \omega_C$ dominates,
the ground state is a spin singlet, with 0$\uparrow$ and 0$\downarrow$ 
Landau levels fully occupied.  On the other hand, when 
$E_Z$ is the largest energy, the ground state is fully polarized; when 
the Coulomb interaction is not strong enough to cause substantial Landau level
mixing (i.e., in the limit of $E_Z>>\hbar \omega_C>>V_C$), the ground state has 
0$\uparrow$ and 1$\uparrow$ Landau levels occupied.  When the ground state is
described in terms of filled Landau levels, we will denote it by $(N$$\uparrow :
N$$\downarrow)$, where $N$$\uparrow$ and $N$$\downarrow$ are the numbers of
occupied Landau levels for up and down spin electrons.  
The possible filled Landau level states at $\nu=2$ are
then $(1:1)$ and $(2:0)$, the unpolarized and the fully polarized states,
respectively.  Our interest will be in situations
when $V_C$ becomes comparable to or greater than the cyclotron energy.
This again corresponds to $r_S\geq 1$, where at
$\nu=2$, $r_S$ can be seen to be given by
\begin{equation}
r_S  = \frac{V_C}{\hbar \omega_C}\;,
\label{eq:r_S}
\end{equation} 
and is clearly a measure of 
the strength of the interaction relative to the cyclotron
energy.  Here, Landau level mixing becomes crucial and may destabilize 
the above states.  Our principal goal will be to determine the phase diagram
of the fully polarized and the unpolarized states $(2:0)$ and $(1:1)$.

Our work has been motivated by the recent experiments of Eriksson {\em
et al.} \cite {Pinczuk} where they investigate by inelastic light
scattering both the spin and 
the charge density collective modes at
$\nu=2$ for samples with $r_S$ as large as 6, corresponding to
densities as low as $\rho = 0.9 \times {10}^{10}$ {cm}$^{-2}$.  
They find a qualitative change in the number
and the character of collective modes at approximately $r_S\approx 2$.   
This was interpreted in a Landau Fermi liquid approach, 
where the magnetic field was treated as a perturbation on the zero
field Fermi liquid. However, the Fermi liquid approach, which
is suitable at small magnetic field, is not an obviously
valid starting point for the problem at hand, and other
approaches are desirable.  A comparison between the ground state energies
of the unpolarized and the fully polarized 
Hartree Fock state shows that
a transition between them 
takes place at $r_S\approx 2.1$ for $E_Z=0$, as we will see in 
\mbox{Sec.\ref{sec-phase}}.  
This raises the question: Is the ground state at large $r_S$ 
fully polarized?  If true, this would indeed be an interesting 
example of an interaction driven ferromagnetism.
If it is indeed fully polarized, is the $(2:0)$ state 
a reasonable starting point for its study?  Besides being fully polarized, 
$(2:0)$ incorporates the effect of Landau level mixing, albeit in a very 
special manner, through promoting all electrons in 0$\downarrow$ to
1$\uparrow$.

A reliable treatment of Landau level mixing lies at the crux
of the problem.   We shall incorporate Landau level mixing in a
perturbative time-dependent Hartree-Fock scheme, i.e., by incorporating 
vertex corrections through ladder diagrams in the  
random phase approximation (RPA).
The most crucial approximation in our calculations 
will be a restriction to the subspace of a single-particle hole pair;
within this subspace, however, the Landau level mixing is treated
accurately.  Clearly, this approach is not quantitatively valid
except at small $r_S$, but we believe that it gives an insight
into the physics even when $r_S$ is not small. 
For a better quantitative description at 
large $r_S$, it would be important to deal with screening 
by more than one particle-hole pair, but that is outside the scope of the
present paper.

\begin{figure}
\centerline{\psfig{figure=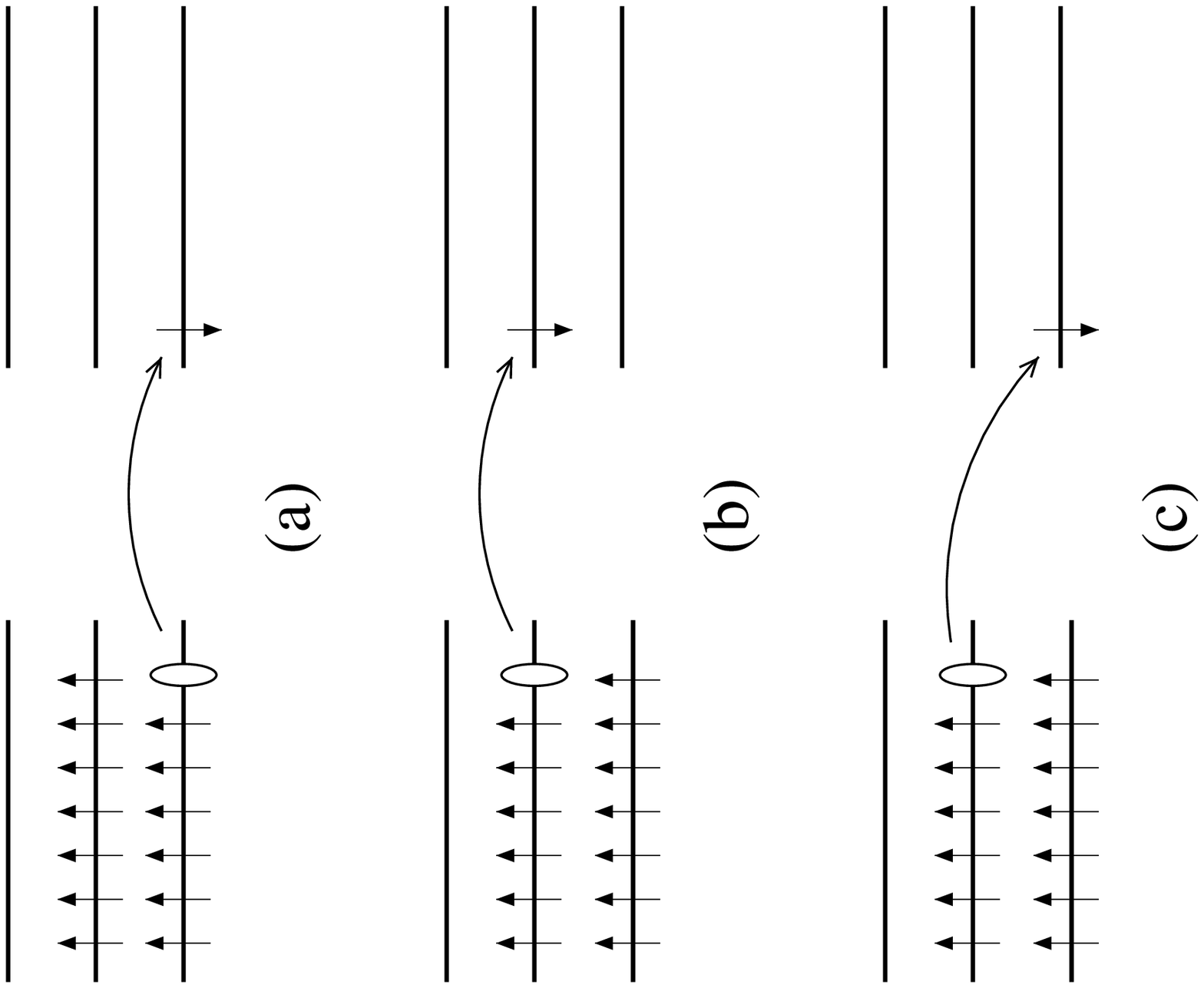,width=7.0in,angle=-90}}
\caption{Schematic diagrams for some relevant excitations contributing to 
the lowest spin density excitations from the 
fully polarized IQHE state at $\nu = 2$. The Zeeman splitting is set to 
zero for simplicity.  The processes of moving
an electron from 0$\uparrow$ to 0$\downarrow$ Landau level
and from 1$\uparrow$ to 1$\downarrow$ Landau level 
are described in figure 
(a) and (b) respectively. Figure (c) depicts the process of
moving an electron from 1$\uparrow$ to 0$\downarrow$ Landau level. 
These are labeled by the kinetic energy change in the small $r_S$ limit
as  $m=0$ (a and b) and $m=-1$ (c) modes. 
\label{sdediagram}}
\end{figure}

As mentioned above, there certainly are parameters for which the  
fully polarized $(2:0)$ state 
describes the actual ground state; in particular, it may occur even when 
$r_S$ is large provided that the Zeeman energy is sufficiently strong.
Our approach here will be to take it as the starting point and 
investigate the regime of its stability by calculating the
dispersion of the charge and the spin density collective modes.
The collective modes have a simple interpretation when 
the Landau level mixing is negligible ($V_C/\hbar \omega_C\rightarrow 0$).
The lowest charge density excitation mode corresponds to the 
excitation of one electron from 1$\uparrow$ to 2$\uparrow$ Landau level.
The kinetic energy change of any collective mode is well defined when the Landau
level mixing is weak, and will be used to label the various modes (this
notation will be used even when the Landau
level mixing is significant, by looking at the evolution of the modes from small
to large $r_S$).
The 1$\uparrow$$\rightarrow$2$\uparrow$ mode 
will be referred to as the $m=1$ mode.
For spin-density excitation, there are three modes of primary 
interest, corresponding
to excitations 
0$\uparrow$$\rightarrow$0$\downarrow$,
1$\uparrow$$\rightarrow$1$\downarrow$, and
1$\uparrow$$\rightarrow$0$\downarrow$ which are
depicted schematically in (a), (b), and (c) in Fig.\ref{sdediagram}
respectively.  The first two are $m=0$ modes and the last one is $m=-1$ mode. 
These are of course coupled, and a diagonalization of the problem produces 
the usual spin-wave excitation mode, with the energy approaching the Zeeman
splitting in the long wave length limit, in accordance with the Goldstone
theorem, as well as two massive spin-density modes. 
Our principal result is that while the charge density collective 
mode shows no instability,  the spin-density
collective mode develops a deep roton minimum in the presence of 
substantial Landau level mixing  
and becomes soft in certain parameter regimes.  
(Indeed, as $r_S$ is increased, 
there will eventually 
also be an instability in the charge density channel, indicating the
formation of a Wigner crystal, but we have not explored this question 
since our 
perturbation theory is not valid at very large $r_S$.)  
Both the existence of the
roton minimum and its softening as the Zeeman energy is reduced are
experimentally testable predictions of our theory.  
The phase diagram thus 
obtained later is shown in Fig.\ref{phase}.  It is instructive
to compare it with the phase diagram in the absence of interactions, which
consists of the fully polarized state
$(2:0)$ for $E_Z>\hbar \omega_C$ and the unpolarized state $(1:1)$ for $E_Z<
\hbar \omega_C$ with a transition taking place at precisely $E_Z=\hbar
\omega_C$. 
At small $r_S$, interactions make the $(2:0)$ state more stable,
indicated by the fact that it survives even for $E_Z< \hbar \omega_C$.  
However,  at large $r_S$, $E_Z>\hbar \omega_C$ is required to
stabilize the fully polarized state.  
It is noteworthy that at small $E_Z$, the 
fully polarized state is found to be unstable at arbitrary $r_S$.
We also consider the collective modes of $(1:1)$, expected to be 
valid at small $E_Z$
and $r_S$.  It becomes unstable as $r_S$ is increased, consistent with
an earlier conclusion of MacDonald \cite{MacDonald2}.

What about the state at large $r_S$ but small Zeeman energy?  As 
discovered in our study, here the $\nu=2$ state is not described by 
either of the two aforementioned Hartree Fock states.  
The finite wave vector spin-wave instability of the fully polarized state
suggests that it is some kind of spin density wave state.  It is not 
possible to be more definitive about it based on our present study. 
Of course, at extremely large $r_S$, when $V_C$ is much larger than the
cyclotron energy, our calculation is unreliable, but indicates
that even if a fully polarized state occurs, as expected based on the zero
field result, it will most likely not be described by 
the $(2:0)$ Hartree Fock state.

There have been many theoretical studies 
of the collective excitations of integer quantum Hall effect (IQHE)
states in the past, but, to our knowledge,
the spin-density wave excitations of the fully polarized $(2:0)$ state 
have not been considered previously.  For other collective modes, our
results reduce to the earlier results in appropriate limits.  If 
possible transitions of the electron and hole between different 
Landau levels are ignored and if self-energy corrections are omitted,
the problem of collective excitation is reduced to determining 
the binding energy of two oppositely charged particles 
strictly confined in their respective  
Landau levels\cite{Binding}. 
In this case the wave function for the bound state is independent of 
interaction potential, and is 
uniquely determined by the wave vector $\vec{k}$.  
The transition of the electron and hole
or the recombination of the particle-hole pair has been considered in the
random phase approximation(RPA)\cite{RPA1,RPA2}.
Later the RPA was incorporated with
the self-energy correction and the binding energy term
by Kallin and Halperin \cite{Halperin} where    
a number of interesting collective excitations from the unpolarized
and the partially polarized ground state
were considered in the absence of 
Landau level mixing (valid when $\hbar \omega_C
>> V_C$).
The Landau level mixing was considered by MacDonald in 
\mbox{Ref \cite{MacDonald2}}, 
treating the mixing matrix elements between various modes
as small parameters and applying a second order perturbation
theory. 
Our calculation will be formulated in terms of diagrams, 
following Kallin and Halperin in \mbox{Ref.\cite{Halperin}}, and
will be performed with a full treatment of the mixing matrix elements
within the subspace of a single 
particle-hole pair at any given instant. 
The diagrammatic formulation of the problem is presented 
in detail in Sec.\ref{sec-diagram} below.

In Sec.\ref{sec-fullypol} we describe the diagrammatic formalism
used to compute dispersion curves of the collective 
excitation from the fully polarized ground state at $\nu=2$. 
We will concentrate on the spin density excitation which will be responsible
for an instability of the IQHE state in the parameter regime under
consideration.  Similarly, 
\mbox{Sec.\ref{sec-unpol}} is devoted to the collective excitations
of the unpolarized ground state at $\nu=2$. Dispersion curves
of the spin density excitation is computed for various values of 
$r_S$ and Zeeman splitting energy, $E_Z$. 
The phase diagram as a function of $r_S$ and $E_Z$ is  obtained in
\mbox{Sec.\ref{sec-phase}} by determining the critical $E_Z$ at which the 
energy of the spin density excitation vanishes. From the phase diagram
we will learn that for large $r_S$ and small $E_Z$ neither the fully 
polarized nor the unpolarized state is stable against a 
spin-density wave state.  The paper is concluded in 
\mbox{Sec.\ref{sec-conclusion}},
where we also discuss the implications of our study for the fractional quantum
Hall effect (FQHE).

\section{Diagrammatic formalism of collective excitation}
\label{sec-diagram}

\subsection{Algebra in the symmetric gauge}
Even though the choice of gauge does not affect 
physical quantities, we find it convenient 
to use the symmetric gauge.  We will start by establishing
the basic algebra in the symmetric gauge 
closely following \mbox{Ref.\cite{MacDonald1}}.  

The Hamiltonian for an electron moving in a two dimensional
space under a perpendicular magnetic field is given by

\begin{equation}
H = \frac{{\vec{\pi}}^2}{2m}
\end{equation}
where the kinetic momentum is written down as

\begin{equation}
\vec{\pi} = -i \hbar \vec{\nabla} + \frac{e \vec{A}}{c}.
\end{equation}

When the magnetic field is uniform,
we can use a convenient algebraic method analogous to the solution
by the ladder operator of the one-dimensional harmonic oscillator. 
In order to construct the algebraic formalism we first
note that the $x$ and $y$ components of the kinetic momentum are
canonically conjugate coordinates: 
 
\begin{equation}
[ \pi_x , \pi_y ] = \frac{-i \hbar e}{c} \hat{z} \cdot (\vec{\nabla} \times \vec{A}) 
= \frac{-i {\hbar}^2 }{ {l_0}^2}.
\end{equation}
From the commutation relationship between $\pi_x$ and $\pi_y$
we can define a ladder operator so that the ladder operator 
and its Hermitian conjugate 
satisfy the same commutation relation as those
of the one-dimensional harmonic oscillator.
That is,
\begin{equation}
[ a , a^{\dagger} ] = 1 ,
\end{equation}
where
\begin{equation}
a^{\dagger} \equiv \frac{ l_0 / \hbar}{\sqrt{2}} ( \pi_x +i \pi_y ).
\end{equation}
The Hamiltonian can now be written in the form of a one-dimensional 
harmonic oscillator.

\begin{equation}
H = \frac{\hbar \omega_C }{2} ( a^{\dagger} a + a a^{\dagger} )
\end{equation}
Therefore the eigenvalues are $\hbar\omega_C(n+1/2)$ 
where $n$ is a non-negative integer 
which is known as the Landau level index.  The eigenstates,
however, cannot be fully determined by the Landau level
index alone because the energy does not depend on the coordinates of 
the cyclotron orbit center, indicating a degeneracy of the Landau level. 
Let us define 
\begin{equation}
\vec{C} = \vec{r} + \frac{ \hat{z} \times \vec{\pi}}{m \omega_C}
\end{equation}
which is convetionally known as the guiding-center operator.
The $x$ and $y$ components of the guiding-center operator are
canonically conjugate coordinates 
similar to those of the kinetic momentum.

\begin{equation}
[ C_x , C_y ] = i {l_0}^2
\end{equation}
Therefore we can define another ladder operator by 
\begin{equation}
b \equiv \frac{1}{\sqrt{2} l_0} ( C_x + i C_y )
\end{equation}
which satisfies  
\begin{equation}
[ b , b^{\dagger} ] = 1 
\end{equation}
and 
\begin{equation}
[ a , b ] = [ a^{\dagger} , b ] = [ H , b ] = 0 .
\end{equation}
The fact that the ladder operator $b$ commutes with the Hamiltonian
shows that the degeneracy of the lowest Landau level 
is actually related to the positioning of the guiding-center
coordinate. Since we identify two independent sets of ladder operators
in a two-dimensional space, the full set of eigenstates can be generated
by repeatedly applying raising operators on the ground state:

\begin{eqnarray}
|n,m \rangle = \frac{ {(a^{\dagger})}^n {(b^{\dagger})}^m }{\sqrt{n! m!}} 
|0,0 \rangle
\end{eqnarray}

For a magnetic field pointing in the positive
$z$ direction, the vector potential in the symmetric gauge is given by

\begin{equation}
\vec{A} = \frac{B}{2} (-y,x,0)
\end{equation}
and the ladder operators may be written in terms of $z (\equiv x +iy)$ as follows.

\begin{equation}
a^{\dagger} = \frac{i}{\sqrt{2}} \Big( \frac{z}{2} -2 \frac{\partial}{\partial \bar{z}} \Big)
\end{equation}

\begin{equation}
a = \frac{-i}{\sqrt{2}} \Big( \frac{\bar{z}}{2} + 2 \frac{\partial}{\partial z} \Big)
\end{equation}

\begin{equation}
b^{\dagger} = \frac{1}{\sqrt{2}} \Big( \frac{\bar{z}}{2} -2 \frac{\partial}{\partial z} \Big)\end{equation}

\begin{equation}
b = \frac{1}{\sqrt{2}} \Big( \frac{z}{2} + 2 \frac{\partial}{\partial \bar{z}} \Big)
\end{equation}
with the Hamiltonian given by  

\begin{equation}
H = \frac{\hbar \omega_C}{2} \Big( -4 \frac{\partial}{\partial z} \frac{\partial}{\partial \bar{z}}+ z\frac{\partial}{\partial z} -\bar{z} \frac{\partial}{\partial \bar{z}} 
+\frac{z\bar{z}}{4} \Big),
\end{equation}
Here and in the rest of the paper, we use the convention 
that $\hbar=c=e=l_0=1$, as well as the 
Area$=1$.  In particular, this implies that  the total number of flux
quanta piercing the system, $N_{\phi}=
\frac{1}{2\pi}$.

Up to this point we have concentrated on the single particle Hamiltonian.
In order to compute concrete physical quantities
with the interaction treated perturbatively, we will need to use certain
matrix elements and their various properties, which we now list \cite
{MacDonald1}.

$\bullet$ \emph{Plane Wave Matrix Elements}

The matrix element of a plane wave $\exp(-i\vec{k}\cdot\vec{r})$ is given by

\begin{equation}
\langle n_{\beta},m_{\beta} |e^{-i \vec{k} \cdot \vec{r}} | n_{\alpha},m_{\alpha} \rangle =
{(-i)}^{n_{\beta}-n_{\alpha}} g_{n_{\beta}n_{\alpha}}(\bar{\kappa})  g_{m_{\beta}m_{\alpha}}(\kappa )e^{- k^2 /2 } ,
\label{eq:planewave}
\end{equation}
where $k=\sqrt{{k_x}^2 +{k_y}^2}$, $\kappa = k_x + i k_y$ and 

\begin{eqnarray}
g_{n_{\beta}n_{\alpha}}(\kappa) 
&\equiv&\langle n_{\beta}|\exp(\frac{-i}{\sqrt{2}} \kappa b^{\dagger})
\exp(\frac{-i}{\sqrt{2}} \bar{\kappa}b)|n_{\alpha}\rangle 
\nonumber\\
&=& {\Big( \frac{2^{n_{\alpha}} n_{\alpha}!}{2^{n_{\beta}} n_{\beta}!} \Big) }^{1/2} 
{ (-i \kappa) }^{n_{\beta}-n_{\alpha}} L^{n_{\beta}-n_{\alpha}}_{n_{\alpha}}
( k^2 /2 )
\label{eq:g}
\end{eqnarray}
where, for $n_{\beta}>n_{\alpha}$, $L^{n_{\beta}-n_{\alpha}}_{n_{\alpha}}$ 
is the associated Laguerre polynomial, defined as:
\begin{equation}
L^m_n(x)=\sum_{s=0}^{n} \frac{(-x)^s}{s!} {{n+m \choose n-s}}
\end{equation} 
For $n_{\beta}<n_{\alpha}$ we define:
\begin{equation}
L^{n_{\beta}-n_{\alpha}}_{n_{\alpha}}(x)=\frac{n_{\beta}!}{n_{\alpha}!} 
(-x)^{n_{\alpha}-n_{\beta}} L^{n_{\alpha}-n_{\beta}}_{n_{\beta}}
\end{equation}
\mbox{Eq.(\ref{eq:planewave})} can be evaluated first by rewriting
$\vec{k}\cdot\vec{r}= (\kappa\bar{z}+\bar{\kappa}z)/2$, expressing
$z$ and $\bar{z}$ in terms of the ladder operators, and moving all
annihilation operators to the right using $e^A e^B=e^B e^A e^{[A,B]}$.

$\bullet$ \emph{Matrix Products}
 
One of the most important properties of the matrix 
$g_{m m'}(\kappa)$ is its product

\begin{equation}
\sum_l g_{n_{\alpha} l}(\kappa_1) g_{l n_{\beta}}(\kappa_2) 
= e^{- \bar{\kappa_1} \kappa_2 /2}  g_{n_{\alpha} n_{\beta}} (\kappa_1+\kappa_2)
\label{eq:product}
\end{equation}
This can be derived by using the definition 
in Eq.~(\ref{eq:g}) and the completeness
of the ladder operator eigenstates.

$\bullet$ \emph{Eigenfunctions in the Symmetric Gauge}

Similar to duality of the position-space and momentum-space
in the one-dimensional harmonic oscillator, 
we have the orbital wave function
closely related to the plane-wave matrix element. 

\begin{equation}
\langle \vec{r} | n,m \rangle \equiv \phi_{n,m}(\vec{r})
= {(-i)}^{n} g_{m n}(i\bar{z}) \frac{e^{-r^2 /4}}{\sqrt{2 \pi}}
\label{eq:phi}
\end{equation}
To derive this, first note that 
where $|0,0 \rangle$ is given by
\begin{equation}
\langle \vec{r}|0,0 \rangle=\frac{1}{\sqrt{2\pi}} \exp[-\frac{1}{4} z\bar{z}]
\end{equation}
which is annihilated by both $a$ and $b$.  The eigenfunction for a general $n$
and $m$ is therefore given by
\begin{equation}
\langle \vec{r} | n,m \rangle=\frac{i^n}{\sqrt{2\pi 2^{m+n}n!m!}} (\frac{z}{2}-2 \frac{\partial}{\partial 
\bar{z}})^n (\frac{\bar{z}}{2}-2 \frac{\partial}{\partial 
z})^m \exp[-\frac{1}{4} z\bar{z}]
\end{equation}
Now write $e^{-z\bar{z}/4}=e^{z\bar{z}/4} e^{-z\bar{z}/2}$ and use
\begin{equation}
\exp[-\frac{1}{4} z\bar{z}]
(\frac{\bar{z}}{2}-2 \frac{\partial}{\partial z})^m
(\frac{z}{2}-2 \frac{\partial}{\partial 
\bar{z}})^n \exp[\frac{1}{4} z\bar{z}]=(-2)^{m+n} (\frac{\partial}{\partial 
\bar{z}})^n (\frac{\partial}{\partial z})^m
\end{equation}
Defining $t=z\bar{z}/2=r^2/2$ one gets
\begin{equation}
\langle \vec{r} | n,m \rangle =
\frac{(-i)^n}{\sqrt{2\pi 2^{n+m}n!m!}} e^{t/2} 2^m z^{n-m} (\frac{\partial}
{\partial t})^n t^m e^{-t}
\end{equation}
which reduces to Eq.~(\ref{eq:phi}) with the standard definition of the
associated Laguerre polynomial.

$\bullet$ \emph{Trace}

The trace of $g_{m m'}(\kappa)$ is a constant because
the charge density of a completely filled Landau level is a constant.

\begin{equation}
\frac{1}{A}\sum^{N_{\phi}}_{m=0} g_{m m}(\kappa) 
= \frac{N_{\phi}}{A} \delta_{\vec{k},0} 
=\frac{1}{2\pi} \delta_{\vec{k},0}
\end{equation}
where $A$ stands for the area of system and therefore
$N_\phi/A$ is $1/2\pi$.

Using the definition of $g_{m m'}(\kappa)$, \mbox{Eq.(\ref{eq:g})},
and properties of the Laguerre polynomial
we find the transpose and complex conjugate
as follows.

$\bullet$ \emph{Transpose}
\begin{equation}
g_{n_{\alpha} n_{\beta}}(\kappa) = g_{n_{\beta} n_{\alpha}}(\bar{\kappa}).
\end{equation}

$\bullet$ \emph{Complex Conjugate}
\begin{equation}
{\bar{g}}_{n_{\alpha} n_{\beta}}(\kappa) = g_{n_{\alpha} n_{\beta}}(-\bar{\kappa})
\end{equation}

$\bullet$ \emph{Orthogonality}

The final property we will use is the orthogonality between
the plane-wave matrix elements which will be 
useful in computing the self-energy:

\begin{equation}
\int d^2 \vec{k} e^{- k^2 /2} g_{n_{\alpha} n_{\beta}}(\kappa)
 g_{{n_{\alpha}}' {n_{\beta}}'}(\bar{\kappa}) = 2 \pi
\delta_{n_{\alpha},{n_{\alpha}}'} \delta_{n_{\beta},{n_{\beta}}'}
\label{eq:orthogonality}.
\end{equation}

\subsection{Response function}

The response function is a very important quantity from which we
can deduce many physical observables assuming linear response.
In particular, the collective excitations correspond
to the poles of the response function.    
In order to calculate the response function
we use the zero temperature limit of the Matsubara formalism\cite{Mahan},
using the standard 
analytic continuation ($i\omega \rightarrow \omega +i\delta$) 
in order to get the retarded response function. We will
describe below in detail only 
the charge-density response function, since the spin-density response
function can be obtained with a straightforward modification.
The charge-density response function is related to the density-density
correlation function as follows:  

\begin{equation}
\chi(k,i\omega) \equiv -\int^{\infty}_{0} d\tau e^{i \omega \tau}
\langle T_{\tau} \hat{\rho}(\vec{k},\tau) \hat{\rho}(-\vec{k},0) \rangle .
\end{equation}
where the density operator, $\hat{\rho}(\vec{k},\tau)$,
is given in the symmetric gauge by 
 
\begin{equation}
\hat{\rho}(\vec{k},\tau) \equiv 
\sum_{n_{\alpha}m_{\alpha}} \sum_{n_{\beta}m_{\beta}} 
\langle n_{\alpha},m_{\alpha}|e^{i\vec{k}\cdot\vec{r}}
|n_{\beta},m_{\beta} \rangle
c^{\dagger}_{n_{\alpha}m_{\alpha}}(\tau)
c_{n_{\beta}m_{\beta}}(\tau).
\end{equation}

\begin{figure}
\centerline{\psfig{figure=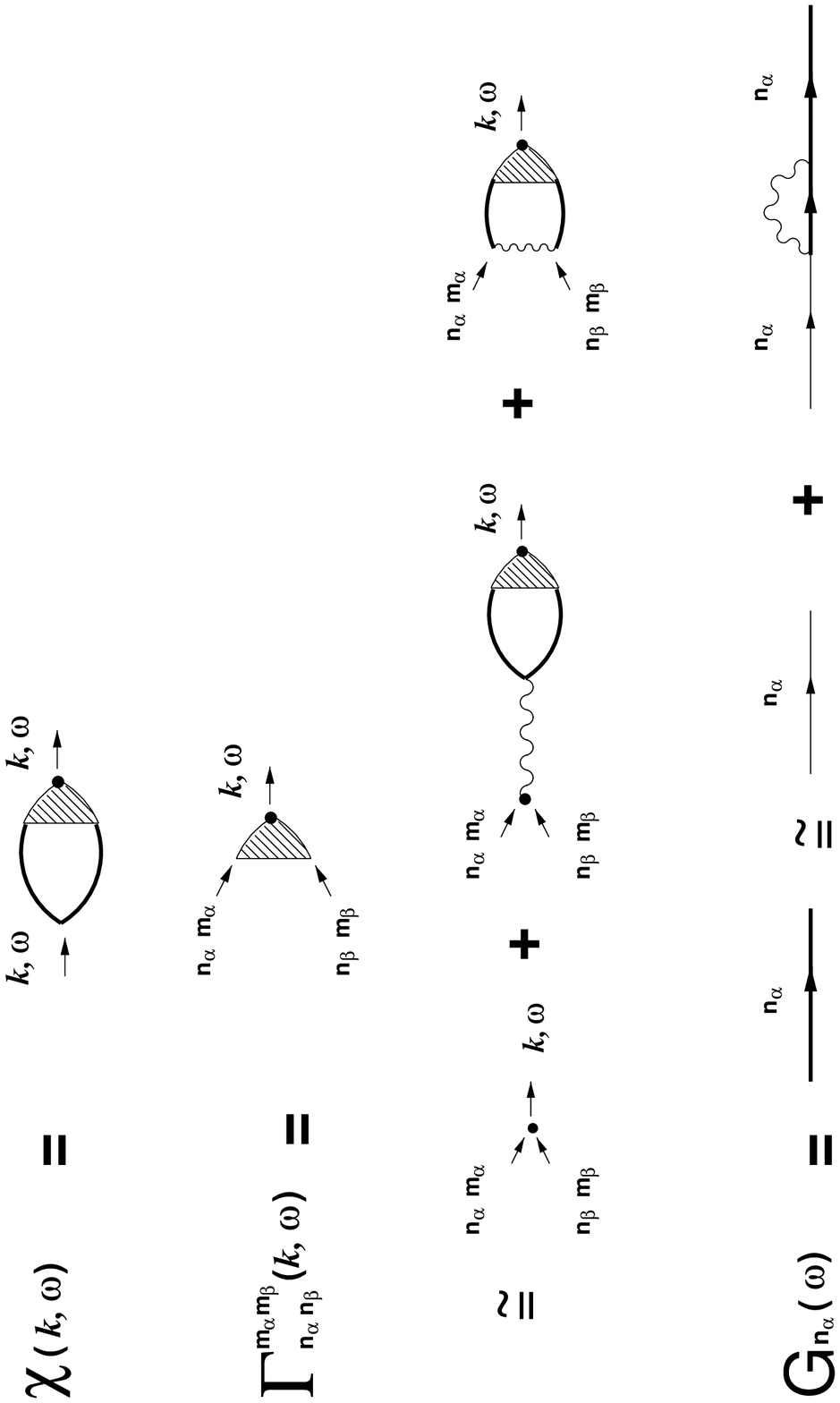,width=7.0in,angle=-90}}
\caption{Feynman diagrams for the response function $\chi(k,\omega)$,
vertex function $\Gamma^{m_\alpha m_\beta}_{n_\alpha n_\beta}(k,\omega)$
and the self-energy correction to 
the ``full'' Green function $G_{n_\alpha}(\omega)$.
\label{vertex}}
\end{figure}

The actual computation of the density response function is performed 
in the perturbation theory. Assuming at any given time 
there is a single particle-hole pair in the system, we write
the response function which is depicted by the Feynman diagram
in \mbox{Fig.\ref{vertex}}.

\begin{equation}
\chi(k,i\omega)= \sum_{n_{\alpha} n_{\beta}}\sum_{m_{\alpha} m_{\beta}}
\langle n_{\alpha},m_{\alpha} |e^{i \vec{k} \cdot \vec{r}} | n_{\beta},m_{\beta} \rangle
D_{n_{\alpha}n_{\beta}}(i\omega)\Gamma^{m_{\alpha} m_{\beta}}_{n_{\alpha} n_{\beta}}(k,i\omega)
\label{eq:chi1}
\end{equation}
where the vertex function 
$\Gamma^{m_{\alpha} m_{\beta}}_{n_{\alpha} n_{\beta}}(k,i\omega)$
satisfies the following
equation:
 
\begin{eqnarray}
\Gamma^{m_{\alpha} m_{\beta}}_{n_{\alpha} n_{\beta}}&(&k,i\omega) =  
\langle n_{\beta},m_{\beta} |e^{-i \vec{k} \cdot \vec{r}} | n_{\alpha},m_{\alpha} \rangle
\nonumber \\
&+&\langle n_{\beta},m_{\beta} |e^{-i \vec{k} \cdot \vec{r}} | n_{\alpha},m_{\alpha} \rangle
\tilde{v}(k) \sum_{{n_{\alpha}}' {n_{\beta}}'}\sum_{{m_{\alpha}}' {m_{\beta}}'}
\langle {n_{\alpha}}',{m_{\alpha}}' |e^{i \vec{k} \cdot \vec{r}} | {n_{\beta}}',{m_{\beta}}' \rangle 
\nonumber \\
&\times& D_{{n_{\alpha}}'{n_{\beta}}'}(i\omega) \Gamma^{m_{{\alpha}'} {m_{\beta}}'}_{{n_{\alpha}}' {n_{\beta}}'}(k,i\omega)
\nonumber \\
&-&\sum_{{n_{\alpha}}' {n_{\beta}}'}\sum_{{m_{\alpha}}' {m_{\beta}}'}
\Bigg[ \int d^2 \vec{r_1}\int d^2 \vec{r_2} \phi_{ n_{\alpha},m_{\alpha} } (\vec{r_1})
{\bar{\phi}}_{ n_{\beta},m_{\beta} } (\vec{r_2}) v(\vec{r_1}-\vec{r_2})
{\bar{\phi}}_{ {n_{\alpha}}',{m_{\alpha}}' } (\vec{r_1})
\phi_{ {n_{\beta}}',{m_{\beta}}' }(\vec{r_2}) \Bigg]
\nonumber \\
&\times& D_{{n_{\alpha}}'{n_{\beta}}'}(i\omega) \Gamma^{m_{{\alpha}}' {m_{\beta}}'}_{{n_{\alpha}}' {n_{\beta}}'}(k,i\omega).
\label{eq:Gamma1}
\end{eqnarray}
where $v(r)$ is a potential and $\hat{v}(k)$ is its Fourier transform.
The Feynman diagrams corresponding to each term 
on the right-hand side of the \mbox{Eq.~(\ref{eq:Gamma1})}
are shown in \mbox{Fig.\ref{vertex}}:
The first term describes just the bare vertex and the second and third
terms depict the RPA correction (Bubble diagram) and the binding 
energy term (Ladder diagram), respectively. 
In the Eq.~(\ref{eq:chi1}) and Eq.~(\ref{eq:Gamma1}) 
$D_{{n_{\alpha}}'{n_{\beta}}'}(i\omega)$ is defined as the
frequency integral of the product of two Green functions:

\begin{eqnarray}
D_{{n_{\alpha}}'{n_{\beta}}'}(i\omega) \equiv
\int \frac{d{\omega}'}{2\pi} G_{{n_{\alpha}}'}(i{\omega}'-i\omega) G_{{n_{\beta}}'}(i{\omega}')
\label{eq:D1}
\end{eqnarray}
and the Green function is given by

\begin{equation}
 G_{n}(i\omega) = \frac{1}{i\omega -(n-\mu_0) \omega_C -\Sigma^{n_0}_n}
\label{eq:Green1} ,
\end{equation}
where $n$ is the Landau level index of the electron or hole 
and $\mu_0$ is the chemical potential which is halfway between the highest
occupied Landau level and the lowest unoccupied Landau level.
We note that the Green function is fully dressed, containing 
the self energy ($\Sigma$) correction, and is also independent of the $m$
quantum number. 
In general the self energy $\Sigma$ is a very complicated function of 
$\vec{k}$ and $\omega$, but it turns out that in this system 
the self energy is real and depends on just the Landau 
level index but not $\vec{k}$ and $\omega$.  We will use this result below, 
postponing the explicit derivation to the next section.

Plugging the explicit form of Green function into Eq.(\ref{eq:D1}) gives

\begin{eqnarray}
D_{{n_{\alpha}}'{n_{\beta}}'}(i\omega) &\equiv&
\int \frac{d{\omega}'}{2\pi} G_{{n_{\alpha}}'}(i{\omega}'-i\omega) G_{{n_{\beta}}'}(i{\omega}')
\nonumber \\
&=& \frac{ \theta\Big( (\mu_0-{n_{\beta}}')\omega_C 
- \Sigma^{n_0}_{{n_{\beta}}'} \Big) - 
\theta\Big( (\mu_0-{n_{\alpha}}')\omega_C 
- \Sigma^{n_0}_{{n_{\alpha}}'} \Big)
}{i\omega - ({n_{\alpha}}'-{n_{\beta}}')\omega_C -(\Sigma^{n_0}_{{n_{\alpha}}'}
-\Sigma^{n_0}_{{n_{\beta}}'}) }
\nonumber \\
&\approx&
 \frac{ \theta(\mu_0-{n_{\beta}}') - 
\theta(\mu_0-{n_{\alpha}}')
}{i\omega - ({n_{\alpha}}'-{n_{\beta}}')\omega_C -(\Sigma^{n_0}_{{n_{\alpha}}'}
-\Sigma^{n_0}_{{n_{\beta}}'}) }
\label{eq:D}
\end{eqnarray}
where 
\begin{equation}
\theta(x) = 
\left\{
\begin{array}{cc}
1 & \mbox{if $x > 0$} \\
0 & \mbox{if $x < 0$} \\
1/2 & \mbox{if $x = 0$}
\end{array}
\right.
\end{equation}
and it is assumed that the self energy does not modify the 
Fermi level significantly but moves the position of the pole
in the denominator of \mbox{Eq.(\ref{eq:D})}.

In order to solve the vertex equation, \mbox{Eq.(\ref{eq:Gamma1})},
we first eliminate the angular momentum index by defining a new vertex
function $\Gamma_{n_{\alpha}n_{\beta}}(k,i\omega)$:

\begin{eqnarray}
\Gamma_{n_{\alpha}n_{\beta}}(k,i\omega) &\equiv&
 e^{-k^2 /2}\sum_{m_{\alpha}m_{\beta}} g_{m_{\alpha}m_{\beta}}(-\kappa)
\Gamma^{m_{\alpha}m_{\beta}}_{n_{\alpha}n_{\beta}}(k,i\omega)
\label{eq:msum1}
\\
&\Updownarrow& \nonumber \\
\Gamma^{m_{\alpha}m_{\beta}}_{n_{\alpha}n_{\beta}}(k,i\omega)
&=& 2\pi{\bar{g}}_{m_{\alpha}m_{\beta}}(-\kappa)\Gamma_{n_{\alpha}n_{\beta}}(k,i\omega).
\label{eq:msum2}
\end{eqnarray}
Eq.(\ref{eq:msum2}) is obtained from Eq.(\ref{eq:msum1}) 
through the property of the plane-wave matrix product 
\mbox{Eq.(\ref{eq:product})}. Noting that the bare vertex is just a
plane-wave matrix element defined in the Eq.(\ref{eq:planewave})
and utilizing the properties of the plane-wave matrix elements, we
succeed in summing over all the angular momentum indices 
to get a new vertex equation.

\begin{eqnarray}
\Gamma_{n_{\alpha} n_{\beta}}&(&k,i\omega) =  
\frac{{(-i)}^{n_{\beta}-n_{\alpha}}}{2\pi} 
e^{-k^2 /2} {\bar{g}}_{n_{\alpha}n_{\beta}}(-\bar{\kappa})
\nonumber \\
&+&
{(-i)}^{n_{\beta}-n_{\alpha}}
e^{-k^2 /2}{\bar{g}}_{n_{\alpha}n_{\beta}}(-\bar{\kappa})
\frac{\tilde{v}(k)}{2\pi} 
\sum_{{n_{\alpha}}' {n_{\beta}}'}{(-i)}^{{n_{\beta}}'-{n_{\alpha}}'}g_{{n_{\alpha}}'{n_{\beta}'}}(-\bar{\kappa})
D_{{n_{\alpha}}'{n_{\beta}}'}(i\omega) \Gamma_{{n_{\alpha}}' {n_{\beta}}'}(k,i\omega)
\nonumber \\
&-&\sum_{{n_{\alpha}}' {n_{\beta}}'}
\Bigg[ 2\pi \int d^2 \vec{r_1}\int d^2 \vec{r_2} \Phi^{\kappa}_{ n_{\alpha},n_{\beta} } (\vec{r_1},\vec{r_2})
v(\vec{r_1}-\vec{r_2})
{\bar{\Phi}}^{\kappa}_{ {n_{\alpha}}',{n_{\beta}}'} (\vec{r_1},\vec{r_2})
\Bigg] D_{{n_{\alpha}}'{n_{\beta}}'}(i\omega) \Gamma_{{n_{\alpha}}' {n_{\beta}}'}(k,i\omega)
\label{eq:Gammasum}
\end{eqnarray}
where
\begin{equation}
\Phi^{\kappa}_{n_{\alpha}n_{\beta}}(\vec{r_1},\vec{r_2}) \equiv
e^{-k^2 /4} \sum_{m_{\alpha} m_{\beta}} g_{m_{\alpha} m_{\beta}}(-\kappa)
\phi_{n_{\alpha} m_{\alpha}}(\vec{r_1}){\bar{\phi}}_{n_{\beta} m_{\beta}}(\vec{r_2}).
\label{eq:Phi}
\end{equation}

Similarly, the density response function can
be written in terms of the new vertex function, 
$\Gamma_{n_\alpha n_\beta}(k,i\omega)$.

\begin{equation}
\chi(k,i\omega)= \sum_{n_{\alpha} n_{\beta}}
{(-i)}^{n_{\alpha}-n_{\beta}} g_{n_{\alpha}n_{\beta}}(-\bar{\kappa})
D_{n_{\alpha}n_{\beta}}(i\omega)\Gamma_{n_{\alpha} n_{\beta}}(k,i\omega)
\label{chi}
\end{equation}
The third term on the right hand side of 
Eq.(\ref{eq:Gammasum}) contains a formal expression for
the Coulomb-potential matrix elements between the wave function 
$\Phi^{\kappa}_{n_{\alpha}n_{\beta}}(\vec{r_1},\vec{r_2})$ and 
$\Phi^{\kappa}_{{n_{\alpha}}'{n_{\beta}}'}(\vec{r_1},\vec{r_2})$
which prove to be the eigenfunctions of the two-body
Hamiltonian for oppositely charged particles confined in the Landau
levels $(n_\alpha,n_\beta)$ and $({n_\alpha}',{n_\beta}')$,
respectively \cite {Halperin}. Therefore, following Kallin and Halperin,
 we will call it the binding energy. The second
term in Eq.(\ref{eq:Gammasum}) has the physical interpretation of  
the exchange energy of particle-hole pair which is the only term
in the usual RPA. For convenience 
let us define $n_{\alpha \beta} =n_{\alpha}-n_{\beta}$, and 
denote the binding energy and the exchange energy as 
$V^{{n_{\alpha}}'{n_{\beta}}'}_{n_{\alpha}n_{\beta}}(k)$ and
$U^{{n_{\alpha}}'{n_{\beta}}'}_{n_{\alpha}n_{\beta}}(k)$
respectively:

\begin{equation}
U^{{n_{\alpha}}'{n_{\beta}}'}_{n_{\alpha}n_{\beta}}(k) \equiv
{i}^{n_{\alpha \beta}-{n_{\alpha \beta}}'}e^{-k^2 /2}{\bar{g}}_{n_{\alpha}n_{\beta}}(-\bar{\kappa}) \frac{\tilde{v}(k)}{2\pi} g_{{n_{\alpha}}'{n_{\beta}'}}(-\bar{\kappa})
\label{eq:U}
\end{equation}

\begin{equation}
V^{{n_{\alpha}}'{n_{\beta}}'}_{n_{\alpha}n_{\beta}}(k) \equiv
2\pi \int d^2 \vec{r_1}\int d^2 \vec{r_2} \Phi^{\kappa}_{ n_{\alpha},n_{\beta} } (\vec{r_1},\vec{r_2})
v(\vec{r_1}-\vec{r_2})
{\bar{\Phi}}^{\kappa}_{ {n_{\alpha}}',{n_{\beta}}'} (\vec{r_1},\vec{r_2})
\label{eq:V}
\end{equation}
Then Eq.~(\ref{eq:Gammasum}) becomes  

\begin{eqnarray}
\Gamma_{n_{\alpha}n_{\beta}}(k,i\omega) &=&
\frac{{i}^{n_{\alpha\beta}}}{2\pi}
e^{-k^2 /2} {\bar{g}}_{n_{\alpha}n_{\beta}}(-\bar{\kappa})
\nonumber \\
&-&
\sum_{{n_{\alpha}}'{n_{\beta}}'} \Big[ V^{{n_{\alpha}}'{n_{\beta}}'}_{n_{\alpha}n_{\beta}}(k)-U^{{n_{\alpha}}'{n_{\beta}}'}_{n_{\alpha}n_{\beta}}(k) \Big]
 D_{{n_{\alpha}}'{n_{\beta}}'}(i\omega) 
\Gamma_{{n_{\alpha}}' {n_{\beta}}'}(k,i\omega).
\end{eqnarray}
The Green function $D_{n_{\alpha},n_{\beta}}$ contains a factor 
$\Big( \theta(n_0 +1/2 -{n_{\beta}})-\theta(n_0 +1/2 -{n_{\alpha}})
\Big)$ which vanishes for certain choices of $(n_{\alpha},n_{\beta})$.
However, these choices co not contribute to $\chi$ in Eq.~(\ref{chi}).
Therefore, restricting to only  
those values of $(n_{\alpha},n_{\beta})$ for
which $D_{n_{\alpha},n_{\beta}}$ is non-zero, we transform 
the equation into an associated set of linear equations:

\begin{eqnarray}
\sum_{{n_{\alpha}}'{n_{\beta}}'} 
\Big[ \delta_{n_{\alpha},{n_{\alpha}}'}
\delta_{n_{\beta},{n_{\beta}}'} D^{-1}_{{n_{\alpha}}'{n_{\beta}}'}(i\omega)
+V^{{n_{\alpha}}'{n_{\beta}}'}_{n_{\alpha}n_{\beta}}(k)&-&U^{{n_{\alpha}}'{n_{\beta}}'}_{n_{\alpha}n_{\beta}}(k) \Big] 
D_{{n_{\alpha}}'{n_{\beta}}'}(i\omega) 
\Gamma_{{n_{\alpha}}' {n_{\beta}}'}(k,i\omega)
\nonumber \\
&=&
\frac{{i}^{n_{\alpha\beta}}}{2\pi}
e^{-k^2 /2} {\bar{g}}_{n_{\alpha}n_{\beta}}(-\bar{\kappa}),
\label{eq:GammaUV}
\end{eqnarray}
where
\begin{equation}
D^{-1}_{{n_{\alpha}}'{n_{\beta}}'}(i\omega) =
\Big( i\omega-({n_{\alpha}}'-{n_{\beta}}')\omega_C
-(\Sigma^{n_0}_{{n_{\alpha}}'}-\Sigma^{n_0}_{{n_{\beta}}'}) \Big)
\times \Big( \theta(n_0 +1/2 -{n_{\beta}}')-\theta(n_0 +1/2 -{n_{\alpha}}')
\Big).
\label{eq:Dinv}
\end{equation}
(Note that here $D^{-1}_{{n_{\alpha}}'{n_{\beta}}'}(i\omega)$ is the
inverse of the matrix element, not the matrix element of the inverse.)
Eq.(\ref{eq:GammaUV}) can be written in the form of a matrix equation
by defining a matrix $M$ as follows:

\begin{equation}
M^{{n_{\alpha}}'{n_{\beta}}'}_{n_{\alpha}n_{\beta}}(k,i\omega) \equiv
 \delta_{n_{\alpha},{n_{\alpha}}'}
\delta_{n_{\beta},{n_{\beta}}'} D^{-1}_{{n_{\alpha}}'{n_{\beta}}'}(i\omega)
+V^{{n_{\alpha}}'{n_{\beta}}'}_{n_{\alpha}n_{\beta}}(k)-U^{{n_{\alpha}}'{n_{\beta}}'}_{n_{\alpha}n_{\beta}}(k)
\label{eq:M} .
\end{equation}
$M^{{n_{\alpha}}'{n_{\beta}}'}_{n_{\alpha}n_{\beta}}(k,i\omega)$ can be
viewed as a matrix element of $M$ if we consider the set of indices 
$(n_{\alpha}n_{\beta})$ to be a collective index. Therefore 
we write Eq.(\ref{eq:GammaUV}) in the following form:

\begin{eqnarray}
\sum_{{n_{\alpha}}'{n_{\beta}}'}
M^{{n_{\alpha}}'{n_{\beta}}'}_{n_{\alpha}n_{\beta}}(k,i\omega)
D_{{n_{\alpha}}'{n_{\beta}}'}(i\omega)
\Gamma_{{n_{\alpha}}' {n_{\beta}}'}(k,i\omega) =
\frac{{i}^{n_{\alpha\beta}}}{2\pi}
e^{-k^2 /2} {\bar{g}}_{n_{\alpha}n_{\beta}}(-\bar{\kappa}).
\label{eq:GammaM}
\end{eqnarray}

Inverting the matrix $M$ amounts to solving the vertex equation.
That is,

\begin{eqnarray}
D_{n_{\alpha}n_{\beta}}(i\omega) \Gamma_{n_{\alpha} n_{\beta}}(k,i\omega) =
\sum_{{n_{\alpha}}'{n_{\beta}}'}
\frac{{i}^{{n_{\alpha\beta}}'}}{2\pi} e^{-k^2 /2}
{(M^{-1})}^{{n_{\alpha}}'{n_{\beta}}'}_{n_{\alpha}n_{\beta}}(k,i\omega)
{\bar{g}}_{{n_{\alpha}}'{n_{\beta}}'}(-\bar{\kappa}).
\end{eqnarray}
where $M^{-1}$ is the inverse matrix of $M$.
Finally, we obtain the density response function in terms of
the matrix $M$. 

\begin{equation}
\chi(k,i\omega)= \sum_{n_{\alpha} n_{\beta}}\sum_{{n_{\alpha}}' {n_{\beta}}'}
\frac{{(-i)}^{n_{\alpha\beta}-{n_{\alpha\beta}}'}}{2\pi} e^{-k^2 /2}
g_{n_{\alpha} n_{\beta}}(-\bar{\kappa})
{(M^{-1})}^{{n_{\alpha}}'{n_{\beta}}'}_{n_{\alpha}n_{\beta}}(k,i\omega)
{\bar{g}}_{{n_{\alpha}}'{n_{\beta}}'}(-\bar{\kappa})
\label{eq:chi}
\end{equation}

Since the summation of the Landau level indices should be performed over all
the possible states, the number of terms is actually infinite. However,
it is possible to obtain an accurate estimate keeping a reasonably 
small number of terms.

\subsection{Collective excitation}
The collective excitations are the poles of the response function, which,
from Eq.(\ref{eq:chi}), correspond to energies for which  
the inverse of $M$ becomes singular, that is to say, when 

\begin{equation}
Det\Big[ M(i\omega \rightarrow \omega + i\delta) \Big] =0.
\label{eq:detM}
\end{equation}

We carry out detailed computations
of the binding energy, RPA energy and self energy 
in order to explicitly evaluate the dispersion curve of the
collective excitation. We start with the binding energy.

$\bullet$ Binding energy (Ladder diagram contribution)

The explicit form of 
$\Phi^{\kappa}_{n_\alpha,n_\beta}(\vec{r_1},\vec{r_2})$
is obtained
from Eq.(\ref{eq:Phi}) by utilizing the plane-wave
matrix product formula, Eq.(\ref{eq:product}): 

\begin{eqnarray}
\Phi^{\kappa}_{ n_{\alpha},n_{\beta} } (\vec{r_1},\vec{r_2})
&=& 
\frac{{(-i)}^{n_{\alpha \beta}}}{2\pi}
e^{-k^2 /4} \sum_{m_{\alpha} m_{\beta}}
g_{m_{\alpha} m_{\beta}} (-\kappa)
e^{-z_1 \bar{z_1} /4} g_{m_{\alpha} n_{\alpha}} (i\bar{z_1})
e^{-z_2 \bar{z_2} /4} {\bar{g}}_{m_{\beta} n_{\beta}} (i\bar{z_2})
\nonumber \\
&=& 
\frac{{(-i)}^{n_{\alpha \beta}}}{2\pi}
e^{-k^2 /4} e^{-({r_1}^2 + {r_2}^2)/4}
\sum_{m_{\alpha} m_{\beta}} 
g_{n_{\alpha} m_{\alpha}} (-i z_1)
g_{m_{\alpha} m_{\beta}} (-\kappa)
g_{m_{\beta} n_{\beta}} (i z_2)
\nonumber \\
&=& 
\frac{{(-i)}^{n_{\alpha \beta}}}{2\pi}
e^{-k^2 /4} e^{-({r_1}^2 + {r_2}^2)/4}
e^{i \bar{\kappa} z_2 /2}
\sum_{m_{\alpha}}
g_{n_{\alpha} m_{\alpha}} (-i z_1)
g_{m_{\alpha} n_{\beta}} (-\kappa+i z_2)
\nonumber \\
&=& 
\frac{{(-i)}^{n_{\alpha \beta}}}{2\pi}
e^{-k^2 /4} e^{-({r_1}^2 + {r_2}^2)/4}
e^{i(\bar{\kappa}z_2 + \kappa\bar{z_1})/2} e^{\bar{z_1} z_2 /2}
g_{n_{\alpha} n_{\beta}} (-\kappa -i z_1 +i z_2)
\nonumber \\
&=&
\frac{{(-i)}^{n_{\alpha \beta}}}{2\pi}
e^{i \vec{R} \cdot (\vec{k} + \vec{r}\times\hat{z}/2)}
e^{-\frac{{| \vec{r}+\vec{k}\times\hat{z} |}^{2}}{4}}
{\bar{g}}_{n_{\alpha} n_{\beta}} ( -i(\bar{z}+i\bar{k}) ).
\end{eqnarray}

Following Kallin and Halperin, the wave function 
$\Phi^{\kappa}_{ n_{\alpha},n_{\beta} } (\vec{r_1},\vec{r_2})$
can be shown to be an eigenstate of the following 
Hamiltonian where 
the particles 1 and 2 are projected onto the Landau levels with
indices $n_{\alpha}$ and $n_{\beta}$ respectively.

\begin{eqnarray}
\hat{H} &=& \hat{P}_{n_\alpha,n_\beta} H \hat{P}_{n_\alpha,n_\beta}
\nonumber \\
&=&
\hat{P}_{n_\alpha,n_\beta} 
\Bigg[ \frac{{\vec{\pi_1}}^{2}}{2m} + \frac{{\vec{\pi_2}}^{2}}{2m}
+v_{arb}(|\vec{r_1}-\vec{r_2}|) \Bigg] 
\hat{P}_{n_\alpha,n_\beta}
\nonumber \\
&=&
\hbar\omega_C (n_\alpha +n_\beta +1) 
+\hat{P}_{n_\alpha,n_\beta}v_{arb}(|\vec{r_1}-\vec{r_2}|)
\hat{P}_{n_\alpha,n_\beta}
\end{eqnarray}
where $\vec{\pi_1}$ $(\vec{\pi_2})$ is the kinetic momentum 
for the particle 1 (2), $v_{arb}(r)$ is an arbitrary (attractive)  
potential and $\hat{P}_{n_\alpha,n_\beta}$ is the operator
projecting the particles onto the Landau levels with
indices $n_{\alpha}$ and $n_{\beta}$. One can prove that 
$\Phi^{\kappa}_{ n_{\alpha},n_{\beta} } (\vec{r_1},\vec{r_2})$
is the eigenfunction of $\hat{H}$
by computing the overlap between 
$v_{arb}(|\vec{r_1}-\vec{r_2}|) \times
\Phi^{\kappa}_{ n_{\alpha},n_{\beta} } (\vec{r_1},\vec{r_2})$
and any arbitrary basis state in the projected Hilbert space, 
for example $\bar{\phi}_{n_\alpha m_\gamma}(\vec{r_1}) 
\phi_{n_\beta m_\delta}(\vec{r_2})$.
Then we note that
it is proportional to the overlap between 
$\Phi^{\kappa}_{ n_{\alpha},n_{\beta} } (\vec{r_1},\vec{r_2})$
and $\bar{\phi}_{n_\alpha m_\gamma}(\vec{r_1}) 
\phi_{n_\beta m_\delta}(\vec{r_2})$. Of course, the proportionality
constant is the eigenvalue which 
is equal to the binding energy previously defined in Eq.(\ref{eq:V})
in the case of ${n_\alpha}'= n_\alpha$ and ${n_\beta}'= n_\beta$.

Now let us get the explicit formula for the binding energy.

\begin{eqnarray}
V^{{n_{\alpha}}'{n_{\beta}}'}_{n_{\alpha}n_{\beta}}(k) 
&=&
2\pi \int d^2 \vec{r_1}\int d^2 \vec{r_2} 
\Phi^{\kappa}_{ n_{\alpha},n_{\beta} } (\vec{r_1},\vec{r_2})
v(\vec{r_1}-\vec{r_2})
{\bar{\Phi}}^{\kappa}_{ {n_{\alpha}}',{n_{\beta}}'} (\vec{r_1},\vec{r_2})
\nonumber \\
&=&
\frac{{(-i)}^{n_{\alpha\beta}-{n_{\alpha \beta}}'}}{2\pi}
\int d^2 \vec{R} \int d^2 \vec{r}
e^{-\frac{{| \vec{r}+\vec{k}\times\hat{z} |}^{2}}{2}}
v(\vec{r})
{\bar{g}}_{n_{\alpha} n_{\beta}} ( -i(\bar{z}+i\bar{k}) )
g_{{n_{\alpha}}' {n_{\beta}}'} ( -i(\bar{z}+i\bar{k}) )
\nonumber \\
&=&
{(-i)}^{n_{\alpha\beta}-{n_{\alpha \beta}}'}
\int \frac{d^2 \vec{r}}{2\pi}
v(\vec{r}-\vec{k}\times\hat{z})
e^{- r^2 /2}
{\bar{g}}_{n_{\alpha} n_{\beta}} ( -i\bar{z})
g_{{n_{\alpha}}' {n_{\beta}}'} ( -i\bar{z})
\nonumber \\
&=&
{(-i)}^{n_{\alpha\beta}-{n_{\alpha \beta}}'}
\int \frac{d^2 \vec{q}}{{(2\pi)}^2} \tilde{v}(q) e^{-i\vec{q}\cdot\vec{k}\times\hat{z}}
\int \frac{d^2 \vec{r}}{2\pi} e^{i\vec{q}\cdot\vec{r}}e^{- r^2 /2}
{\bar{g}}_{n_{\alpha} n_{\beta}} ( -i\bar{z})
g_{{n_{\alpha}}' {n_{\beta}}'} ( -i\bar{z})
\nonumber \\
&=&
{(-i)}^{n_{\alpha\beta}-{n_{\alpha \beta}}'}
{\Bigg( \frac{2^{n_{\beta}}2^{{n_{\beta}}'} n_{\beta}! {n_{\beta}}'!}
{2^{n_{\alpha}}2^{{n_{\alpha}}'} n_{\alpha}! {n_{\alpha}}'!} \Bigg)}^{1/2}
\int \frac{d^2 \vec{q}}{{(2\pi)}^2} \tilde{v}(q) e^{i\vec{q}\cdot\vec{k}\times\hat{z}}
\nonumber \\
&\times& \int \frac{d^2 \vec{r}}{2\pi} e^{i\vec{q}\cdot\vec{r}}e^{- r^2 /2}
z^{n_{\alpha\beta}} {\bar{z}}^{n_{\alpha\beta}'}
L^{n_{\alpha\beta}}_{n_{\beta}}(\frac{r^2}{2}) 
L^{{n_{\alpha\beta}}'}_{{n_{\beta}}'}(\frac{r^2}{2}),
\end{eqnarray}
where $L^m_n$ is an associated Laguerre polynomial. By using the fact 
$\int^{2\pi}_{0} d\theta e^{i(x \cos\theta+n\theta)}= 
2 \pi i^n J_{n}(x)$ 
we perform the angle integrations to get the final formula:

\begin{eqnarray}
V^{{n_{\alpha}}'{n_{\beta}}'}_{n_{\alpha}n_{\beta}}(k) &=&
e^{i(n_{\alpha\beta}-{n_{\alpha \beta}}') \theta_{\vec{k}} }
{\Bigg( \frac{2^{n_{\beta}}2^{{n_{\beta}}'} n_{\beta}! {n_{\beta}}'!}
{2^{n_{\alpha}}2^{{n_{\alpha}}'} n_{\alpha}! {n_{\alpha}}'!} \Bigg)}^{1/2}
\int^{\infty}_{0} dq \frac{q}{2\pi} \tilde{v}(q) 
J_{n_{\alpha\beta}-{n_{\alpha \beta}}'}(qk)
\nonumber \\
&\times&
\int^{\infty}_{0} dr e^{-r^2 /2} r^{n_{\alpha\beta}+{n_{\alpha \beta}}'+1}
L^{n_{\alpha\beta}}_{n_{\beta}}(\frac{r^2}{2}) 
L^{{n_{\alpha\beta}}'}_{{n_{\beta}}'}(\frac{r^2}{2})
J_{n_{\alpha\beta}-{n_{\alpha \beta}}'}(qr),
\label{eq:explicitV}
\end{eqnarray}
where $n_{\alpha\beta}= n_\alpha-n_\beta$,  
${n_{\alpha\beta}}'= {n_\alpha}'-{n_\beta}'$
and $J_n$ is a Bessel function and $\theta_{\vec{k}}$ is the angle of
$\vec{k}$ measured from the $x$-axis.

$\bullet$ Random phase approximation energy (Bubble diagram contribution)

The exchange energy from the RPA is rather straightforward to calculate
since there is no integration involved. The explicit form is given by

\begin{eqnarray}
U^{{n_{\alpha}}'{n_{\beta}}'}_{n_{\alpha}n_{\beta}}(k) &=&
{i}^{n_{\alpha \beta}-{n_{\alpha \beta}}'}e^{-k^2 /2}{\bar{g}}_{n_{\alpha}n_{\beta}}(-\bar{\kappa}) \frac{\tilde{v}(k)}{2\pi} g_{{n_{\alpha}}'{n_{\beta}'}}(-\bar{\kappa})
\nonumber \\
&=&
e^{i(n_{\alpha\beta}-{n_{\alpha \beta}}')\theta_{\vec{k}} }
{\Bigg( \frac{2^{n_{\beta}}2^{{n_{\beta}}'} n_{\beta}! {n_{\beta}}'!}
{2^{n_{\alpha}}2^{{n_{\alpha}}'} n_{\alpha}! {n_{\alpha}}'!} \Bigg)}^{1/2}
\frac{\tilde{v}(k)}{2\pi} e^{-k^2 /2}
k^{n_{\alpha\beta}+{n_{\alpha \beta}}'}
L^{n_{\alpha\beta}}_{n_{\beta}}(\frac{k^2}{2}) 
L^{{n_{\alpha\beta}}'}_{{n_{\beta}}'}(\frac{k^2}{2}).
\label{eq:explicitU}
\end{eqnarray}

Incidentally,
a comparison between Eq.(\ref{eq:explicitV}) 
and Eq.(\ref{eq:explicitU}) reveals that
$U^{{n_{\alpha}}'{n_{\beta}}'}_{n_{\alpha}n_{\beta}}(k)$ and
$V^{{n_{\alpha}}'{n_{\beta}}'}_{n_{\alpha}n_{\beta}}(k)$ have
the same phase factor, 
$e^{i(n_{\alpha\beta}-{n_{\alpha \beta}}')\theta_{\vec{k}}}$.
Therefore the phase factor can be eliminated in a consistent way,
which is expected because the system is uniform and isotropic.

$\bullet$ Self energy

The last diagram in the Fig.\ref{vertex} is
the self-energy contribution to the ``full'' Green function.
As before, a single particle-hole pair is assumed, which is reflected
in the Feynman diagram through the unscreened Coulomb line.
The corresponding Dyson equation is solved in the conventional way:

\begin{eqnarray}
G_{n_{\alpha}}(i\omega) &=& G^{(0)}_{n_{\alpha}}(i\omega)
+ G^{(0)}_{n_{\alpha}}(i\omega) \Sigma^{n_0}_{n_{\alpha}} 
G_{n_{\alpha}}(i\omega)
\nonumber \\
&=& 
\frac{1}{i\omega -(n_{\alpha}-\mu_0)\omega_C -\Sigma^{n_0}_{n_{\alpha}} }
\end{eqnarray}
where

\begin{equation}
G^{(0)}_{n_{\alpha}}(i\omega) = \frac{1}{i\omega -(n_{\alpha}-\mu_0)\omega_C}
\end{equation}
and

\begin{eqnarray}
\Sigma^{n_0}_{n_{\alpha}} &\cong& -\sum_{n_{\beta}m_{\beta}} \Bigg[
\int d^2 \vec{r_1} \int d^2 \vec{r_2}
{\bar{\phi}}_{n_{\alpha}m_{\alpha}}(\vec{r_1})
\phi_{n_{\alpha}m_{\alpha}}(\vec{r_2})
v(\vec{r_1}-\vec{r_2})
\phi_{n_{\beta}m_{\beta}}(\vec{r_1})
{\bar{\phi}}_{n_{\beta}m_{\beta}}(\vec{r_2}) \Bigg]
\nonumber \\
&\times&
\int \frac{d \omega}{2\pi} G^{(0)}_{n_{\beta}}(i\omega)
\nonumber \\
&=&
-\sum_{n_{\beta}m_{\beta}} \Bigg[
\int d^2 \vec{r_1} \int d^2 \vec{r_2}
{\bar{\phi}}_{n_{\alpha}m_{\alpha}}(\vec{r_1})
\phi_{n_{\alpha}m_{\alpha}}(\vec{r_2})
v(\vec{r_1}-\vec{r_2})
\phi_{n_{\beta}m_{\beta}}(\vec{r_1})
{\bar{\phi}}_{n_{\beta}m_{\beta}}(\vec{r_2}) \Bigg]
\nonumber \\
&\times&
\theta( n_0 +1/2 -n_{\beta} )
\nonumber \\
&=&
-\sum^{n_0}_{n_{\beta}=0} \Bigg[
\int d^2 \vec{r_1} \int d^2 \vec{r_2}
{\bar{\phi}}_{n_{\alpha}m_{\alpha}}(\vec{r_1})
\phi_{n_{\alpha}m_{\alpha}}(\vec{r_2})
v(\vec{r_1}-\vec{r_2})
\sum_{m_{\beta}}
\phi_{n_{\beta}m_{\beta}}(\vec{r_1})
{\bar{\phi}}_{n_{\beta}m_{\beta}}(\vec{r_2}) \Bigg]
\nonumber \\
&=&
-\int \frac{d^2 \vec{r_1}}{2\pi} 
\int \frac{d^2 \vec{r_2}}{2\pi}
e^{-({r_1}^2+{r_2}^2 -z_2 \bar{z_1})/2}
v(\vec{r_1}-\vec{r_2})
{\bar{g}}_{m_{\alpha}n_{\alpha}}(i\bar{z_1})
g_{m_{\alpha}n_{\alpha}}(i\bar{z_2})
\sum^{n_0}_{n_{\beta}=0} 
g_{n_{\beta}n_{\beta}}(i(\bar{z_1}-\bar{z_2}))
\end{eqnarray}

In the above equation the explicit form of the single particle eigenstate,
Eq.(\ref{eq:phi}), and the plane-wave matrix product formula,
Eq.(\ref{eq:product}), have been used.
It is convenient for the computation of integrals to change the variables  
from $\vec{r_1}$ and
$\vec{r_2}$ to the center of mass coordinate $\vec{R}$ and the relative 
coordinate $\vec{r}$. In the form of a complex number 
$\vec{R} \leftrightarrow Z \equiv (z_1 + z_2)/2$
and $\vec{r} \leftrightarrow z \equiv z_1 - z_2$.

\begin{eqnarray}
\Sigma^{n_0}_{n_{\alpha}} &=&
-\int \frac{d^2 \vec{r}}{2\pi} v(\vec{r}) e^{-3 r^2 /8}
\sum^{n_0}_{n_{\beta}=0} L^{0}_{n_{\beta}} (\frac{r^2}{2})
\int \frac{d^2 \vec{R}}{2\pi}
e^{- R^2 /2} e^{-(z\bar{Z}-\bar{z}Z)/4}
{\bar{g}}_{m_{\alpha}n_{\alpha}}(i\bar{Z}+\frac{i}{2}\bar{z})
g_{m_{\alpha}n_{\alpha}}(i\bar{Z}-\frac{i}{2}\bar{z})
\nonumber \\
&=&
-\int \frac{d^2 \vec{r}}{2\pi} v(\vec{r}) e^{-r^2 /2}
\sum^{n_0}_{n_{\beta}=0} L^{0}_{n_{\beta}} (\frac{r^2}{2})
\int \frac{d^2 \vec{R}}{2\pi} 
\exp \Big( -\frac{1}{2} {|Z+z/2|}^2 +\frac{1}{2} (Z+z/2)\bar{z} \Big)
\nonumber \\
&\times&
{\bar{g}}_{m_{\alpha}n_{\alpha}}(i\bar{Z}+\frac{i}{2}\bar{z})
g_{m_{\alpha}n_{\alpha}}(i\bar{Z}-\frac{i}{2}\bar{z})
\nonumber \\
&=&
-\int \frac{d^2 \vec{r}}{2\pi} v(\vec{r}) e^{-r^2 /2}
\sum^{n_0}_{n_{\beta}=0} L^{0}_{n_{\beta}} (\frac{r^2}{2})
\int \frac{d^2 {\vec{R}}' }{2\pi} 
\exp \Big( -\frac{1}{2} {|Z'|}^2 \Big)
\nonumber \\
&\times&
{\bar{g}}_{m_{\alpha}n_{\alpha}}(i{\bar{Z}}')
\exp \Big( \frac{1}{2}Z' \bar{z} \Big)
g_{m_{\alpha}n_{\alpha}}(i{\bar{Z}}'-i\bar{z})
\nonumber \\
&=&
-\int \frac{d^2 \vec{r}}{2\pi} v(\vec{r}) e^{-r^2 /2}
\sum^{n_0}_{n_{\beta}=0} L^{0}_{n_{\beta}} (\frac{r^2}{2})
\int \frac{d^2 {\vec{R}}' }{2\pi} 
\exp \Big( -\frac{1}{2} {|Z'|}^2 \Big)
\nonumber \\
&\times&
g_{n_{\alpha}m_{\alpha}}(-i{\bar{Z}}')
\sum_{l} g_{m_{\alpha} l}(i{\bar{Z}}')
g_{l n_{\alpha}}(-i\bar{z})
\nonumber \\
&=&
-\int \frac{d^2 \vec{r}}{2\pi} v(\vec{r}) e^{-r^2 /2}
\sum^{n_0}_{n_{\beta}=0} L^{0}_{n_{\beta}} (\frac{r^2}{2})
\sum_{l} g_{l n_{\alpha}}(-i\bar{z})
\underbrace{
\int \frac{d^2 {\vec{R}}' }{2\pi} \exp \Big( -\frac{1}{2} {|Z'|}^2 \Big)
g_{n_{\alpha}m_{\alpha}}(-i{\bar{Z}}')
g_{l m_{\alpha}}(i Z') }_{= \delta_{l,n_{\alpha}} \mbox{ : orthogonality} }
\nonumber \\
&=&
-\int \frac{d^2 \vec{r}}{2\pi} v(\vec{r}) e^{-r^2 /2}
g_{n_{\alpha} n_{\alpha}}(-i\bar{z})
\sum^{n_0}_{n_{\beta}=0} L^{0}_{n_{\beta}} (\frac{r^2}{2})
\end{eqnarray}

The angular momentum index $m_\alpha$ is naturally eliminated after
the integration over the center-of-mass coordinate by using
Eq.(\ref{eq:orthogonality}). The plane-wave matrix
product formula has also been used when we proceed from the third step to the
fourth step. 
Using $\sum^{n_0}_{n=0} L^{0}_{n}(x) = L^{1}_{n_0}(x)$ one can finally
write down the self energy as follows:

\begin{eqnarray}
\Sigma^{n_0}_{n_{\alpha}} &=& 
-\int \frac{d^2 \vec{r}}{2\pi} v(\vec{r}) e^{-r^2 /2}
L^{0}_{n_{\alpha}} ( \frac{r^2}{2} )
L^{1}_{n_0} (\frac{r^2}{2} )
\nonumber \\
&=&
-\int^{\infty}_{0} dq \frac{q}{2\pi} \tilde{v}(q)
\int^{\infty}_{0} dr \cdot r L^{0}_{n_{\alpha}}(\frac{r^2}{2})
 L^{1}_{n_0}(\frac{r^2}{2}) J_{0}(qr) e^{- r^2 /2}.
\label{eq:Sigma}
\end{eqnarray}

\subsection{Spin degree of freedom}
In order to see what modifications need to  be made in the 
above analysis  to include the spin degree of freedom, it 
is instructive to recall the physical meaning of
the three parts of the collective exciton energy.
The binding energy due to the ladder diagram is the 
\emph{direct} interaction between the excited electron and 
the hole. Therefore it will not be affected by the presence of
the spin degree of freedom. On the other hand, the RPA energy
is the \emph{exchange} energy between the excited electron
and the hole. So it will vanish if the excited electron has a 
different spin than the hole, as in the case of 
the spin density excitation. In other words the electron-hole
pair with the same spin cannot recombine through the Coulomb potential.
In the case of the charge density excitation, however, the RPA energy
depends on the polarization of the ground state. We have computed the
RPA energy in the previous section assuming that all the electrons have
the same spin, which corresponds to the fully polarized state.
If the ground state is unpolarized, the RPA energy will be twice
as large as  that of the fully polarized state simply because
the particle-hole pair can be created and annihilated as either 
a spin-up or spin-down pair. Formally speaking, the vertex
equation, \mbox{Eq.(\ref{eq:Gamma1})}, will have two identical RPA terms for
a given set of indices. 
The self energy term is due to the exchange energy
between a given electron and the
rest of electrons in the system while there is no direct term because we
assume a neutralizing positive charge background. Since it is an
exchange term, we have to include only the interaction between 
the electrons with the same spin. Therefore \mbox{Eq.(\ref{eq:Sigma})}
is generalized to include the spin degree of freedom as follows:

\begin{equation}
\Sigma^{n(\sigma)}_{n} = 
-\int^{\infty}_{0} dq \frac{q}{2\pi} \tilde{v}(q)
\int^{\infty}_{0} dr \cdot r L^{0}_{n}(\frac{r^2}{2})
 L^{1}_{ n(\sigma) }(\frac{r^2}{2}) J_{0}(qr) e^{- r^2 /2}.
\end{equation}
where
\begin{equation}
\sigma = \left\{
\begin{array}{cc}
1/2  &  \mbox{for the spin antiparallel to B field} \\
-1/2 &  \mbox{for the spin parallel to B field}
\end{array}
\right.
\end{equation}
and $n(\sigma)$ is the index of the highest Landau level
occupied by the electron with spin $\sigma$. Using this
new self energy one can rewrite 
\mbox{ Eq.(\ref{eq:D}) } as follows:

\begin{equation}
D_{n_{\alpha} n_{\beta}}(i\omega) 
= \frac{ \theta(\mu(\sigma_\beta)-n_{\beta}) 
- \theta(\mu(\sigma_\alpha)-n_{\alpha})}
{i\omega - (n_{\alpha}-n_{\beta})\omega_C 
-(\sigma_\alpha-\sigma_\beta)E_Z
-(\Sigma^{n(\sigma_\alpha)}_{n_{\alpha}}
-\Sigma^{n(\sigma_\beta)}_{n_{\beta}}) },
\end{equation}
where the chemical potential $\mu(\sigma)$ was defined earlier, and 
the Zeeman coupling is included through the term $\sigma E_z$.

\subsection{Solutions for the pole of the response function}
Computing the dispersion curve of collective
excitations for a general $r_S$ 
requires solving the equation for the pole of the 
response function, \mbox{Eq.(\ref{eq:detM})}. 
In the limit $r_S\rightarrow 0$, when there is no Landau level mixing, 
the matrix has a block diagonal form since collective modes with different 
kinetic energies do not couple, and each block, which has a finite
dimension, can be diagonalized separately to obtain the collective 
mode energies \cite{Halperin}.  
At non-zero values of $r_S$, however, the full matrix
must be diagonalized.  Strictly speaking, the matrix $M$ in 
\mbox{Eq.(\ref{eq:detM})}
is of infinite dimension, but in practice, we work with a finite size matrix,
keeping a sufficient number of Landau levels to ensure a convergence of the 
collective mode energy.  For the lowest energy collective modes,
which are our primary concern, 
and for $r_S<6$, we find that it is adequate to work with $M$ of dimension 
of up to 20.

We also find it useful to convert the Eq.~(\ref{eq:detM}) into an eigenvalue 
equation, the solution of which can be obtained using standard 
linear algebraic methods.  In this formulation
it is natural to define an effective Hamiltonian to be
the matrix in the eigenvalue problem. 
The details of the procedure are discussed in the remainder of this section.

For convenience we write down
\mbox{Eq.(\ref{eq:chi})} here after the analytic continuation. 
\begin{equation}
\chi(k,\omega)= \sum_{n_{\alpha} n_{\beta}}\sum_{{n_{\alpha}}' {n_{\beta}}'}
\frac{{(-i)}^{n_{\alpha\beta}-{n_{\alpha\beta}}'}}{2\pi} e^{-k^2 /2}
g_{n_{\alpha} n_{\beta}}(-\bar{\kappa})
{(M^{-1})}^{{n_{\alpha}}'{n_{\beta}}'}_{n_{\alpha}n_{\beta}}(k,\omega)
{\bar{g}}_{{n_{\alpha}}'{n_{\beta}}'}(-\bar{\kappa}).
\label{eq:chiapdx}
\end{equation}
For an arbitrary set
of indices $(n_\alpha,n_\beta)$ where $n_\alpha$ is greater than 
$n_\beta$, there is a reversed set $(n_\beta,n_\alpha)$ whose 
kinetic energy cost is negative. The mode described by the reversed set
of Landau level indices
was called a negative-energy mode by MacDonald because of its negative 
kinetic energy cost.
\cite{MacDonald2}.  MacDonald
considered mixing between positive and negative energy modes 
as well as between the positive energy modes in second-order
perturbation theory in order to compute the collective 
excitations at general $r_S$ in the spin unpolarized ground state. 
In the present work
we directly solve the pole equation, \mbox{Eq.(\ref{eq:detM})}, 
instead of approximating it to the second-order. 
In computing the lowest lying mode 
it is especially important 
to consider mixing with the 
negative-energy modes since the lowest positive-energy
mode is energetically closest to the negative-energy modes.

In order to make our discussion concrete and transparent, 
let us explicitly write down the matrix elements in the case
of the full spin polarization. If ${n_\alpha}' - {n_\beta}' 
\equiv m > 0$, ${n_\alpha}' > \mu_0$ and ${n_\beta}' < \mu_0$,
\mbox{Eq.(\ref{eq:Dinv})} is written, after the analytic 
continuation, as follows:

\begin{equation}
D^{-1}_{{n_{\alpha}}'{n_{\beta}}'}(\omega) =
\Big( \omega-m\omega_C
-(\Sigma^{n_0}_{{n_{\alpha}}'}-\Sigma^{n_0}_{{n_{\beta}}'}) \Big).
\end{equation}
If the order of the indices is reversed, we get  

\begin{eqnarray}
D^{-1}_{{n_{\beta}}'{n_{\alpha}}'}(\omega) &=&
\Big( \omega+m\omega_C
+(\Sigma^{n_0}_{{n_{\alpha}}'}-\Sigma^{n_0}_{{n_{\beta}}'}) \Big)
\times (-1)
\nonumber \\
&=&
D^{-1}_{{n_{\alpha}}'{n_{\beta}}'}(-\omega)
\end{eqnarray}
Therefore the matrix $M$ defined in
\mbox{Eq.(\ref{eq:M})} and its counterpart with
reversed indices are

\begin{equation}
M^{{n_{\alpha}}'{n_{\beta}}'}_{n_{\alpha}n_{\beta}}(k,\omega) =
 \delta_{n_{\alpha},{n_{\alpha}}'}
\delta_{n_{\beta},{n_{\beta}}'} D^{-1}_{{n_{\alpha}}'{n_{\beta}}'}(\omega)
+V^{{n_{\alpha}}'{n_{\beta}}'}_{n_{\alpha}n_{\beta}}(k)-U^{{n_{\alpha}}'{n_{\beta
}}'}_{n_{\alpha}n_{\beta}}(k)
\label{eq:posM}
\end{equation}
and

\begin{equation}
M^{{n_{\beta}}'{n_{\alpha}}'}_{n_{\beta}n_{\alpha}}(k,\omega) =
 \delta_{n_{\alpha},{n_{\alpha}}'}
\delta_{n_{\beta},{n_{\beta}}'} D^{-1}_{{n_{\alpha}}'{n_{\beta}}'}(-\omega)
+(-1)^{n_{\alpha\beta}-{n_{\alpha\beta}}'}
\Big(
V^{{n_{\alpha}}'{n_{\beta}}'}_{n_{\alpha}n_{\beta}}(k)-U^{{n_{\alpha}}'{n_{\beta}
}'}_{n_{\alpha}n_{\beta}}(k)
\Big)
\label{eq:negM}
\end{equation}
where 
we set $\theta_{\vec{k}} = 0$ without loss of generality and
therefore
$V^{{n_{\beta}}'{n_{\alpha}}'}_{n_{\beta}n_{\alpha}}
( U^{{n_{\beta}}'{n_{\alpha}}'}_{n_{\beta}n_{\alpha}} )
= (-1)^{n_{\alpha\beta}-{n_{\alpha\beta}}'}
V^{{n_{\alpha}}'{n_{\beta}}'}_{n_{\alpha}n_{\beta}}
( U^{{n_{\alpha}}'{n_{\beta}}'}_{n_{\alpha}n_{\beta}} )$.

The sign in front of the second term in \mbox{Eq.(\ref{eq:negM})} can
be interpreted so that the negative-energy mode has the angle of
wave vector equal to $\pi$: for the negative-energy mode
$\theta_{\vec{k}} = \pi$ in the phase factor of 
$V^{{n_{\beta}}'{n_{\alpha}}'}_{n_{\beta}n_{\alpha}}$ and
$U^{{n_{\beta}}'{n_{\alpha}}'}_{n_{\beta}n_{\alpha}}$. 
According to \mbox{Eq.(\ref{eq:negM})} 
the negative-energy mode has a negative frequency and the opposite
direction of wave vector relative to the positive-energy mode.
Therefore when the positive-energy mode is viewed as a plane wave
$e^{i(\omega t-\vec{k}\cdot\vec{r})}$,
the negative-energy mode is the complex conjugate plane wave 
$e^{-i(\omega t-\vec{k}\cdot\vec{r})}$.
In this interpretation the requirement of negative-energy mode is natural
since an arbitrary plane wave with $\vec{k}$ is written as a linear 
combination of $e^{i(\omega t-\vec{k}\cdot\vec{r})}$
and $e^{-i(\omega t-\vec{k}\cdot\vec{r})}$. Therefore 
in general a collective excitation with $\vec{k}$ should
be described by both the positive-energy and negative-energy
modes.  Incidentally, we mention that the mode describing
an excitation within the same Landau level, for example 
the spin-wave Goldstone mode, does not have the negative-energy counterpart
because there should not be any double counting in 
\mbox{Eq.(\ref{eq:chiapdx})}.

In any case we realize that the pole equation, \mbox{Eq.(\ref{eq:detM})}, 
is not an eigenvalue equation as it stands because
of the sign change in $\omega$ for the negative-energy mode. But 
it can be transformed to an eigenvalue equation 
as follows. Let us denote $M$ in terms of sub-matrices.

\begin{equation}
M(k,\omega) =
\left[
\begin{array}{cc}
M_{00}(k,\omega)  &  M_{01}(k) \\
M_{10}(k)         &  M_{11}(k,\omega) 
\end{array}
\right]
\end{equation} 
where $M_{00} (M_{11})$ is associated with mixing between 
the positive-energy (negative-energy) modes
and $M_{01} (= M_{10})$ is between the positive and negative energy modes.
Thanks to a property of determinant we can obtain the solution
of $Det[M(k,\omega)] = 0$ by solving the following equation.

\begin{equation}
Det[\tilde{M} (k,\omega)] = 0 ,
\label{eq:detMtilde}
\end{equation}
where
\begin{equation}
\tilde{M}(k,\omega) =
\left[
\begin{array}{cc}
M_{00}(k,\omega)  &  M_{01}(k) \\
-M_{10}(k)         &  -M_{11}(k,\omega) 
\end{array}
\right]
\end{equation} 
 
Now one can define \emph{an effective Hamiltonian matrix} $H(k)$
using $\tilde{M}(k,\omega)$.

\begin{equation}
\tilde{M}(k,\omega) = H(k) -\omega I
\label{eq:effH}
\end{equation}
Therefore \mbox{Eq.(\ref{eq:detMtilde})} amounts to the eigenvalue
equation of the effective Hamiltonian matrix $H$, which, however, is not
a Hermitian matrix because of the sign change.

Solving the eigenvalue equation of a non-Hermitian
matrix is complicated by the fact that the eigenvalues
of a non-Hermitian matrix can be highly sensitive to small 
changes in the matrix elements \cite{NumRecipes}. The sensitivity
of eigenvalues to rounding errors during the execution of 
some algorithms can be reduced by the procedure of \emph{balancing}.
The idea of balancing is to use similarity transformations to
make corresponding rows and columns of a matrix have comparable
norms, thus reducing the overall norm of the matrix while leaving
the eigenvalues unchanged. Then the general strategy for finding
the eigenvalues of a matrix is to reduce the matrix to a simpler
form, and perform an iterative procedure on the simplified
matrix. The simpler structure we use is called the Hessenberg form.
An upper Hessenberg matrix has zeros everywhere below
the diagonal except for the first subdiagonal row. Then
one can find the eigenvalues by applying the \emph{QR algorithm}
repeatedly to the Hessenberg form until convergence is reached.

Finally, we mention that even though the above formalism has been developed for
a general situation, we will concentrate on the physics
at $\nu=2$ in the following sections taking the Coulomb potential
as the interaction. When the finite thickness effect of the
2D system is of interest, one can replace the Coulomb potential
by an effective potential such as the Stern-Howard potential
\cite{Kallin,thickness}. 

\section{Collective excitations of the fully polarized IQHE state at $\nu =2$}
\label{sec-fullypol}

Equipped with the explicit formulas for the binding energy, 
the RPA energy and the self energy, 
we compute the dispersion curves of the collective excitation from
the fully polarized ground state. First we study the large B field
limit, i.e. the small $r_S$ limit, where the (time-dependent)
Hatree-Fock approximation is valid. 
In the small $r_S$ limit the effective Hamiltonian $H$ 
defined in \mbox{Eq. (\ref{eq:effH})}
is already block-diagonalized so that 
only the matrix elements within the Hilbert subspace
of the same kinetic energy survive.
The off-diagonal terms
due to the interaction energy
become negligible compared to the kinetic energy. Therefore
the pole equation is simple to solve in this case.
For an arbitrary value of
$r_S$ the Hamiltonian is generally complicated and needs to be
diagonalized over whole Hilbert space.

When we consider the small $r_S$ limit,
the integrals we encounter in 
Eq.(\ref{eq:explicitV}), Eq.(\ref{eq:explicitU}) and 
Eq.(\ref{eq:Sigma}) can be expressed in terms of 
$f_n(\alpha,\beta)$ and $g_n(\alpha)$,
defined as follows:

\begin{equation}
f_{n}(\alpha,\beta) \equiv \int^{\infty}_{0} d x \cdot x^{n} J_0 (\beta x)
e^{-\alpha x^2}
\end{equation}
and 

\begin{equation}
g_{n}(\alpha) \equiv \int^{\infty}_{0} d x \cdot x^{n} 
e^{-\alpha x^2}.
\end{equation}
where $n$ is an integer. The explicit functional forms 
of $f_n(\alpha,\beta)$ and $g_n(\alpha)$
are categorized into ones with even $n$ or 
with odd $n$:

\begin{eqnarray}
f_{2m}(\alpha,\beta) &=& {\Bigg( - \frac{\partial}{\partial \alpha} \Bigg)}^m
f_0(\alpha,\beta) 
\nonumber \\
f_{2m+1}(\alpha,\beta) &=& {\Bigg( - \frac{\partial}{\partial \alpha} \Bigg)}^m
f_1(\alpha,\beta) 
\end{eqnarray}
where
\begin{eqnarray} 
f_0(\alpha,\beta) &=& \sqrt{\frac{\pi}{4\alpha}} e^{- {\beta}^2 /8\alpha}
I_0(\frac{{\beta}^2}{8\alpha}),
\nonumber \\
f_1(\alpha,\beta) &=& \frac{1}{2\alpha} e^{- {\beta}^2 /4\alpha}
\end{eqnarray}
and $I_0$ is a modified Bessel function. Similarly,

\begin{eqnarray}
g_{2m}(\alpha) &=& {\Bigg( - \frac{\partial}{\partial \alpha} \Bigg)}^m
g_0(\alpha) 
\nonumber \\
g_{2m+1}(\alpha) &=& {\Bigg( - \frac{\partial}{\partial \alpha} \Bigg)}^m
g_1(\alpha)
\end{eqnarray}
where
\begin{eqnarray}
g_0(\alpha) &=& \sqrt{\frac{\pi}{4\alpha}}, 
\nonumber \\
g_1(\alpha) &=& \frac{1}{2\alpha} 
\end{eqnarray}

\subsection{Charge density excitations}

Since the lowest mode in the energy spectrum is most relevant,
we consider the mode where an electron is taken from the $(n=1)$
Landau level and promoted to the $(n=2)$ Landau level without
flipping the spin in order to obtain a charge density excitation. 
Since its kinetic energy is $\hbar\omega$, we will call the mode
the $m=1$ mode. As discussed before, this mode
is not mixed with other modes in the limit of small $r_S$, i.e., the 
matrix $M$ is already diagonal.  For the lowest charge density
excitation we have the following equation:

\begin{eqnarray}
M^{2,1}_{2,1}(k,\omega) =
\omega -\omega_C
-\frac{e^2}{\epsilon l_0}(\tilde{\Sigma^{1}}_{n=2}
-\tilde{\Sigma^{1}}_{n=1})
-\frac{e^2}{\epsilon l_0} \tilde{U}^{2,1}_{2,1}(k)
+\frac{e^2}{\epsilon l_0} \tilde{V}^{2,1}_{2,1}(k)
=0.
\end{eqnarray}
In the above equation we defined dimensionless matrix elements so that

\begin{eqnarray}
\tilde{V}^{{n_{\alpha}}'{n_{\beta}}'}_{n_{\alpha}n_{\beta}}(k)
&=&
V^{{n_{\alpha}}'{n_{\beta}}'}_{n_{\alpha}n_{\beta}}(k)
/(e^2 /\epsilon l_0)
\nonumber \\\
\tilde{U}^{{n_{\alpha}}'{n_{\beta}}'}_{n_{\alpha}n_{\beta}}(k)
&=&
U^{{n_{\alpha}}'{n_{\beta}}'}_{n_{\alpha}n_{\beta}}(k)
/(e^2 /\epsilon l_0)
\nonumber \\
\tilde{\Sigma}^{n_0}_{n_\alpha}
&=&
\Sigma^{n_0}_{n_\alpha}/(e^2 /\epsilon l_0)
\end{eqnarray}

Let us denote the solution of the pole equation 
as $\Delta(k)$ from now on. Then the dispersion curve of
the lowest charge density excitation 
is given by 

\begin{eqnarray}
\Delta(k) &=&
\omega_C
+\frac{e^2}{\epsilon l_0}(\tilde{\Sigma}^{1}_{n=2}
-\tilde{\Sigma}_{n=1})
-\frac{e^2}{\epsilon l_0} \tilde{V}^{2,1}_{2,1}(k)
+\frac{e^2}{\epsilon l_0} \tilde{U}^{2,1}_{2,1}(k)
\nonumber \\
&=&
\omega_C
+\frac{e^2}{\epsilon l_0}
\Bigg[
g_2(\alpha) -\frac{1}{2}g_4(\alpha) +\frac{1}{16}g_6(\alpha)
\nonumber \\
&-& f_0(\alpha,\beta) +\frac{3}{2}f_2(\alpha,\beta)
-\frac{5}{8}f_4(\alpha,\beta)+\frac{1}{16}f_6(\alpha,\beta)
\Bigg]_{\alpha=1/2,\beta=k}
\nonumber \\
&+& \frac{e^2}{\epsilon l_0} e^{-k^2/2}k {\Bigg( 1-\frac{k^2}{4} \Bigg)}^2
\end{eqnarray}

\begin{figure}
\centerline{\psfig{figure=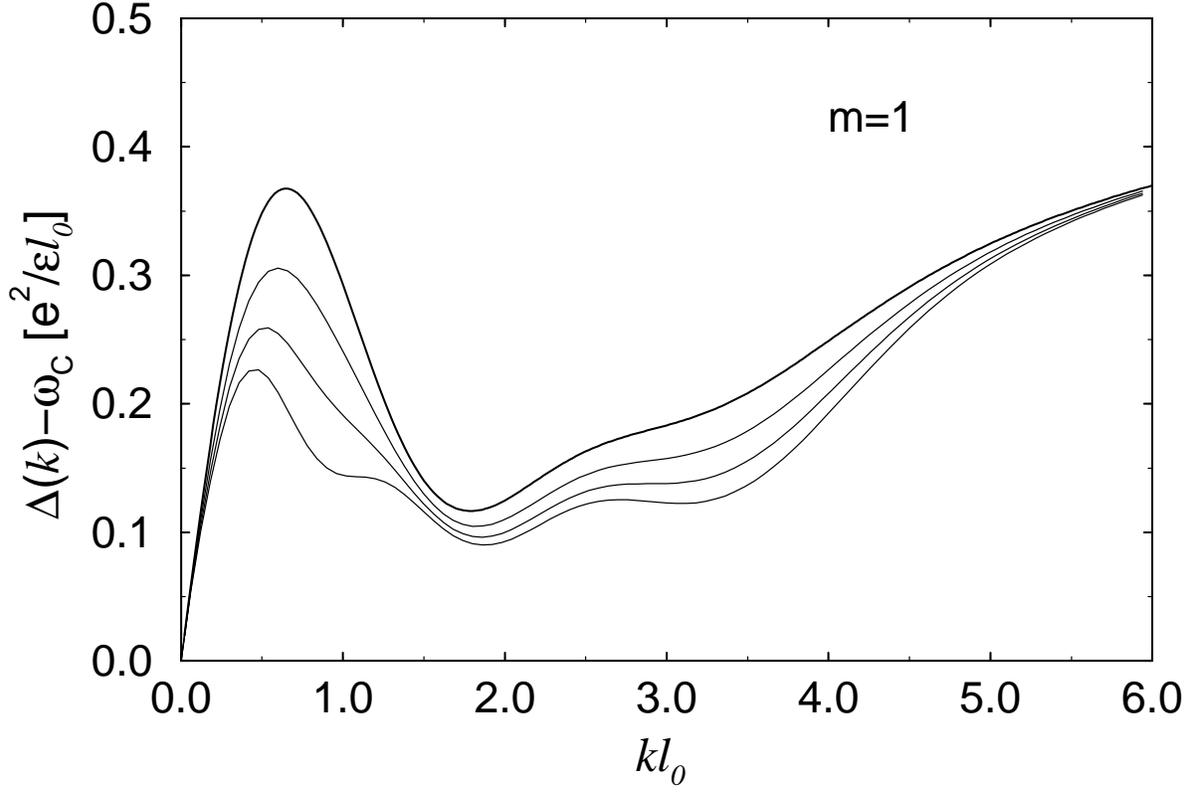,width=7.0in,angle=-90}}
\caption{ Dispersion curves of the lowest charge density excitation 
from the fully polarized IQHE state at $\nu = 2$ for various values of $r_S$. 
>From the top,
the values of $r_S$ are 0.0, 1.0, 2.0 and 3.0. 
We denote this mode by $m=1$ charge density mode, since its kinetic energy
approaches unity (in units of $\hbar \omega_C$) in the limit of  
$r_S = 0.0$ there is no Landau level
mixing.
\label{polcde}}
\end{figure}

As explained earlier,
Landau level mixing in the non-zero $r_S$ regime is included
by diagonalizing the effective Hamiltonian defined in Eq.(\ref{eq:effH}).
Using the term $m=1$ mode for the sake of convenience
to indicate
the lowest charge density excitation
in the fully polarized state at a general $r_S$, 
we plot their dispersion curves in Fig.\ref{polcde}
which shows that the charge density excitation modes do not
exhibit any sign of an instability in the 
parameter range of $r_S$ considered here.  
A non-trivial check of our calculations is to make sure that 
the collective excitations
computed within our approximation satisfy the exact 
Kohn's theorem which states that the $m=1$ mode energy 
must approach $\omega_C$ as $k \rightarrow 0$ \cite{Kohn}.
Fig.\ref{polcde} shows that Kohn's theorem
is satisfied not only for the pure mode but also the mode
with Landau level mixing in general $r_S$. 
Incidentally,
the Zeeman coupling will not affect the dispersion curves
because the spin is not flipped.

\subsection{Spin density excitations}

Following the convention used in the previous sections,
we indicate the excitation modes in terms of
$m$, the kinetic energy of the mode in units of $\hbar \omega_C$ 
in the limit of small $r_S$.  We shall see that, unlike the 
charge density excitation, the lowest lying spin excitation can be either 
$m=-1$ mode or the lower one of the two $m=0$ modes, 
depending on the value
of $r_S$. The $m=-1$ mode describes the process whereby
an electron in the $n=1$ Landau level is demoted
to the $n=0$ Landau level with its spin reversed, whereas 
the $m=0$ mode has an electron with its spin flipped 
in the same Landau level. In the latter case, 
there are two possible modes:
spin density excitation within $n=1$ Landau level or $n=0$
Landau level.  At small $r_S$ the $m=-1$ mode the lowest excitation
while for large $r_S$ the lowest spin density 
excitation is a $m=0$ mode.
Also, without the Zeeman coupling, the spin reversed mode always causes
an instability of the fully polarized (2:0) state.  A determination of the 
Zeeman splitting energy $E_Z$ required to make the (2:0) state  
stable for general $r_S$ will be one of the main goals when we
try to obtain the phase diagram of the spin polarization as
a function of $r_S$ and $E_Z$. The dispersion curves of 
the pure spin density excitation,
however, can be computed in the limit of small $r_S$
without recourse to the actual value of $E_Z$.
We assume that it is large enough to stablize the excitation
because the Zeeman energy is just a constant shift in this limit.

$\bullet$ $m = -1$ mode

\begin{figure}
\centerline{\psfig{figure=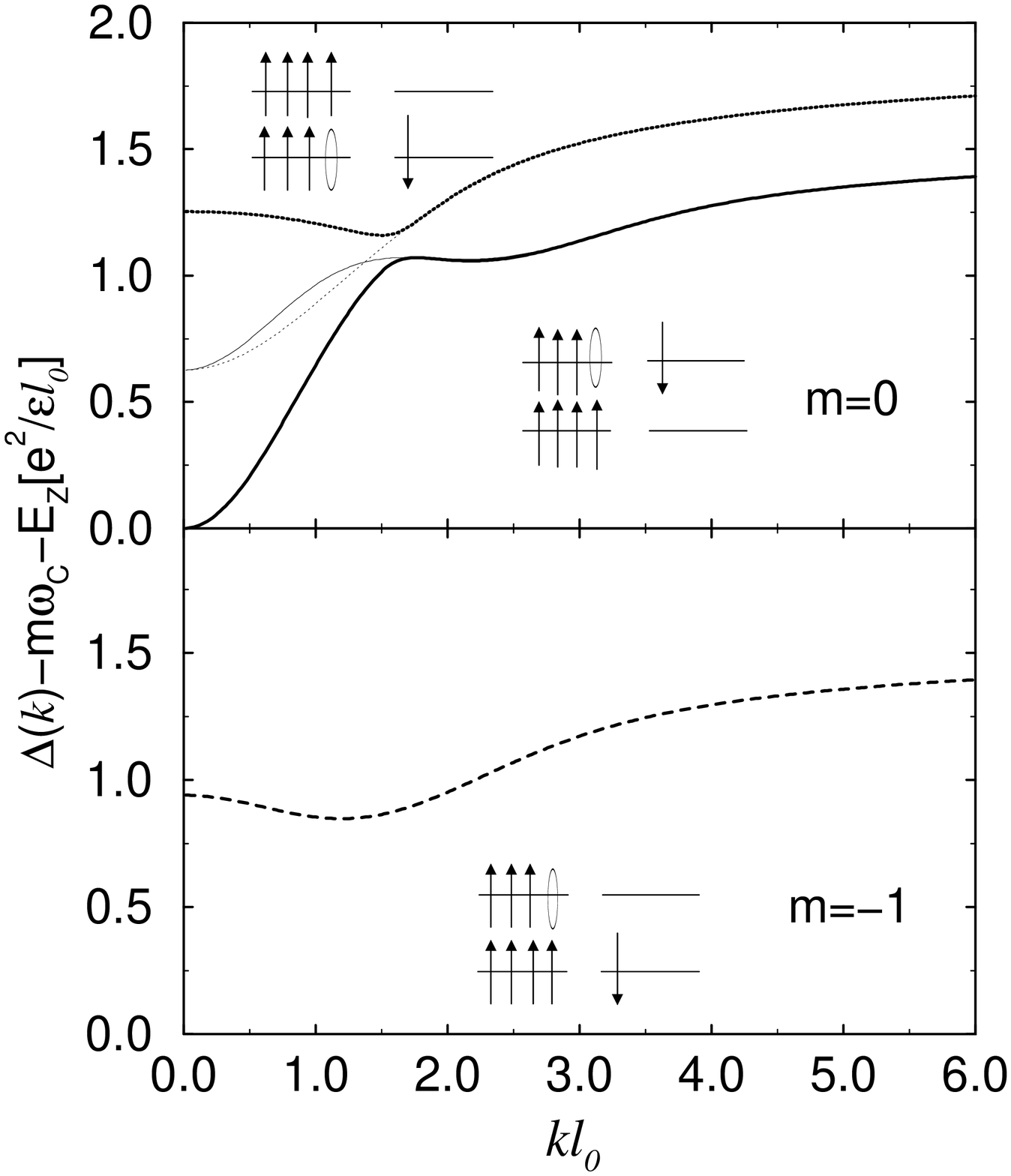,width=5.0in,angle=0}}
\caption{ Dispersion curves of the spin density excitations
in the fully polarized IQHE state at $\nu = 2$ for $r_S = 0$.
Note that when $r_S = 0$ the Zeeman energy contribution is
a constant shift in energy.
Each mode of the spin density excitation is denoted by $m$ i.e.
its kinetic energy in units of $\hbar\omega_C$.
Since there are two degenerate modes for $m=0$, we diagonalize
the Hamiltonian in this Hilbert subspace to get the eigenstates
whose dispersion curves are plotted 
as the thick solid and dotted lines in the top
graph. The thin lines in the top graph indicate the uncoupled modes described
in the adjacent diagrams where $\uparrow$ ($\downarrow$) denotes
the spin-up (-down) electron and the horizontal lines denote the 
Landau levels. The lower mode in the two $m=0$ modes is referred to  
either as the Goldstone mode or the spin-wave excitation mode. 
The dispersion curve of $m = -1$ mode is plotted in the bottom graph.
\label{puresde}}
\end{figure}

As in the case of the charge density excitation, 
we first solve the pole equation in the small $r_S$ limit
to get the pure mode without
any Landau level mixing.
With matrix elements 
\begin{equation}
M^{0,1}_{0,1}(k,\omega) = \omega +\omega_C
- \frac{e^2}{\epsilon l_0}
\Big[ - \tilde{\Sigma}^{1}_{n=1}
-\tilde{V}^{0,1}_{0,1} (k) \Big] = 0
\end{equation}
the solution is

\begin{equation}
\Delta(k) = -\omega_C + \frac{e^2}{\epsilon l_0}
\Big[ - \tilde{\Sigma}^{1}_{n=1}
-\tilde{V}^{0,1}_{0,1} (k) \Big]
\end{equation}
where 

\begin{eqnarray}
- \tilde{\Sigma}^{1}_{n=1} =
\Big[ 2g_0(\alpha) -\frac{3}{2} g_2(\alpha) +\frac{1}{4} g_4(\alpha)
\Big]_{\alpha=1/2}
= \frac{5}{4}   \sqrt{ \frac{\pi}{2}}
\end{eqnarray}
and

\begin{eqnarray}
-\tilde{V}^{0,1}_{0,1} (k) &=& \Big[
-f_0(\alpha,\beta) +\frac{1}{2} f_2(\alpha,\beta) \Big]_{\alpha=1/2,\beta=k}
\nonumber \\
&=&
-\frac{1}{2} \sqrt{  \frac{\pi}{2}} e^{-k^2 /4}
\Bigg[ (1+\frac{k^2}{2})I_0(\frac{k^2}{4}) -\frac{k^2}{2}I_1(\frac{k^2}{4})
\Bigg] 
\end{eqnarray}
The dispersion curve of this $m=-1$ mode is plotted in Fig.\ref{puresde}.

$\bullet$ $m=0$ modes and the spin wave mode

Since there are two possible $m=0$ modes, the pole equation 
becomes a matrix equation even in the small $r_S$ limit:
 
\begin{equation}
\left|
\begin{array}{cc}
M^{0,0}_{0,0}(k,\omega) & M^{0,0}_{1,1}(k) \\
M^{1,1}_{0,0}(k)        & M^{1,1}_{1,1}(k,\omega) 
\end{array} \right|
=0     
\label{eq:detm0}
\end{equation}
The matrix elements are given by

\begin{eqnarray}
M^{0,0}_{0,0}(k,\omega) &=&
\omega - \frac{e^2}{\epsilon l_0} 
\Big[
-\tilde{\Sigma}^{1}_{n=0}
-\tilde{V}^{0,0}_{0,0}(k)
\Big]
\nonumber \\
&=&
\omega- \frac{e^2}{\epsilon l_0} 
\Big[
2g_0(\alpha)-\frac{1}{2}g_2(\alpha)
-f_0(\alpha,\beta)
\Big]_{\alpha=1/2,\beta=k}
\nonumber \\
&=&
\omega - \frac{e^2}{\epsilon l_0} \sqrt{\frac{\pi}{2}}\Bigg[
\frac{3}{2} 
- e^{-k^2 /4} I_0(\frac{k^2}{4}) \Bigg] ,
\\
M^{0,0}_{1,1}(k) &=&
\frac{e^2}{\epsilon l_0} 
\tilde{V}^{0,0}_{1,1}(k)
\nonumber \\
&=&
\frac{e^2}{\epsilon l_0} 
\Big[
\frac{1}{2} f_2(\alpha,\beta)
\Big]_{\alpha=1/2,\beta=k}
\nonumber \\
&=&
\frac{e^2}{\epsilon l_0} \frac{1}{2} \sqrt{\frac{\pi}{2}}
e^{-k^2 /4} \Bigg[
(1-\frac{k^2}{2})I_0(\frac{k^2}{4})+\frac{k^2}{2}I_1(\frac{k^2}{4})
\Bigg] 
\nonumber \\
&=& M^{1,1}_{0,0}(k),
\\
\mbox{and}
\nonumber \\
M^{1,1}_{1,1}(k,\omega) &=&
\omega 
-\frac{e^2}{\epsilon l_0} 
\Big[
-\tilde{\Sigma}^{1}_{n=1}
-\tilde{V}^{1,1}_{0,0}(k)
\Big]
\nonumber \\
&=&
\omega
-\frac{e^2}{\epsilon l_0} 
\Big[
2g_0(\alpha) -\frac{3}{2} g_2(\alpha) +\frac{1}{4}g_4(\alpha)
-f_0(\alpha,\beta)+f_2(\alpha,\beta) -\frac{1}{4}f_4(\alpha,\beta)
\Big]_{\alpha=1/2,\beta=k}
\nonumber \\
&=&
\omega
-\frac{e^2}{\epsilon l_0} 
\frac{5}{4}\sqrt{ \frac{\pi}{2}}
\nonumber \\
&+&
\frac{e^2}{\epsilon l_0}\frac{1}{4} \sqrt{ \frac{\pi}{2}} e^{-k^2 /4}
\Bigg[
(3-k^2+\frac{3}{8}k^4)I_0(\frac{k^2}{4})
+k^2(1-\frac{k^2}{2})I_1(\frac{k^2}{4})
+\frac{k^4}{8}I_2(\frac{k^2}{4})
\Bigg]
\end{eqnarray}

Since spontaneous symmetry breaking occurs in the 
fully polarized state, there must be a spin-wave Goldstone mode whose
energy approaches 
the unshifted Zeeman splitting energy in the long wavelength
limit. Similar to the case of the charge density excitation
it is important to check if 
the spin density excitations computed within our approximation
satisfy this exact theorem. 
When we take the $k \rightarrow 0$ limit of \mbox{Eq.(\ref{eq:detm0})},
we have the following equation:
 
\begin{equation}
\left|
\begin{array}{cc}
\omega -\frac{e^2}{\epsilon l_0} \frac{1}{2}\sqrt{\frac{\pi}{2}}
& 
\frac{e^2}{\epsilon l_0} \frac{1}{2}\sqrt{\frac{\pi}{2}}
\\
\frac{e^2}{\epsilon l_0} \frac{1}{2}\sqrt{\frac{\pi}{2}}
& 
\omega -\frac{e^2}{\epsilon l_0} \frac{1}{2}\sqrt{\frac{\pi}{2}}
\end{array} \right|
=0 .     
\end{equation}
The solutions are 
\begin{eqnarray}
\Delta_1(k=0) &=& 0 \\
\Delta_2(k=0) &=& \frac{e^2}{\epsilon l_0} \sqrt{\frac{\pi}{2}}.
\end{eqnarray}
This confirms the existence of a Goldstone (spin-wave) mode.
Furthermore it can be shown that the
massless spin-wave mode exists 
with an arbitrary potential and Landau level
mixing within our approximation.
The dispersion curves of the two $m=0$ modes are also
plotted in Fig.\ref{puresde} along with that of $m=-1$ mode.

As before, the dispersion curves for general $r_S$ and $E_Z$ are
obtained  from diagonalization of the effective Hamiltonian.
We take the dispersion curves at $r_S =$ 1.0 and 3.0 as examples
and plot them in Fig.\ref{polsde} 
to illustrate the qualitative difference 
between the small and large $r_S$ regimes.  
We will sometimes use the term Goldstone mode in order to
indicate spin-wave excitation mode for general $r_S$ since 
its energy approaches $E_Z$ as $k$ goes to zero.
We will also use the term $m=-1$ mode for the lowest excitation
for small $r_S$ since it has the lowest energy in the limit of
vanishing $r_S$. The physical implications and the corresponding
phase diagram will be discussed in more details in the later section.

\section{Collective excitations from the unpolarized IQHE state at $\nu =2$}
\label{sec-unpol}

The collective excitations in the spin unpolarized state are
computed in this section in complete analogy with the previous section.
First, let us consider the small $r_S$ limit with zero $E_Z$. 
The energy of the lowest charge density excitation is 
\begin{eqnarray}
\Delta(k) &=& \omega_C
+\frac{e^2}{\epsilon l_0} [ 
\tilde{\Sigma}^{0}_{n=1} - 
\tilde{\Sigma}^{0}_{n=0} ]
-\frac{e^2}{\epsilon l_0} \tilde{V}^{1,0}_{1,0}(k)
+\frac{e^2}{\epsilon l_0} 2 \tilde{U}^{1,0}_{1,0}(k)
\nonumber \\
&=& \omega_C +\frac{e^2}{\epsilon l_0}k e^{-k^2 /4}
+\frac{e^2}{\epsilon l_0} \Bigg[
g_0(\alpha) -\frac{1}{2}g_2(\alpha)
-f_0(\alpha,\beta) +\frac{1}{2}f_2(\alpha,\beta) \Bigg]_{\alpha=1/2,\beta=k} .
\end{eqnarray}
And the energy of the lowest spin density exciation is

\begin{eqnarray}
\Delta(k) &=& \omega_C
+\frac{e^2}{\epsilon l_0} [ 
\tilde{\Sigma}^{0}_{n=1} - 
\tilde{\Sigma}^{0}_{n=0} ]
-\frac{e^2}{\epsilon l_0} \tilde{V}^{1,0}_{1,0}(k)
\nonumber \\
&=& \omega_C 
+\frac{e^2}{\epsilon l_0} \Bigg[
g_0(\alpha) -\frac{1}{2}g_2(\alpha)
-f_0(\alpha,\beta) +\frac{1}{2}f_2(\alpha,\beta) \Bigg]_{\alpha=1/2,\beta=k}. 
\end{eqnarray}

\begin{figure}
\centerline{\psfig{figure=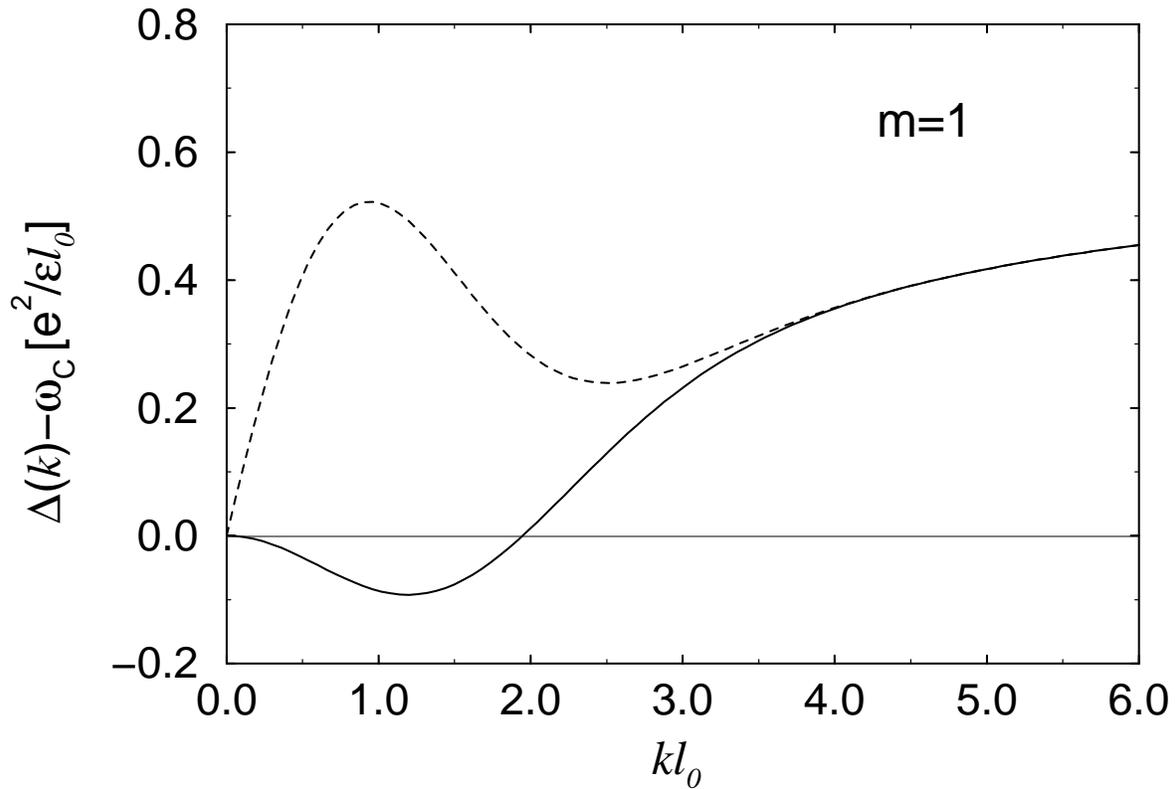,width=7.0in,angle=-90}}
\caption{ Dispersion curves of the lowest charge (dashed line) 
and spin (solid line) density excitation
in the unpolarized IQHE state at $\nu =2$ at $r_S =0$. The Zeeman energy has
been set to zero in the plot;  
for $r_S = 0$ the Zeeman energy contribution to the spin density
excitation dispersion is a constant shift of $E_Z$. The charge
density excitation is not affected by the Zeeman coupling.  
\label{pureun}}
\end{figure}

\begin{figure}
\centerline{\psfig{figure=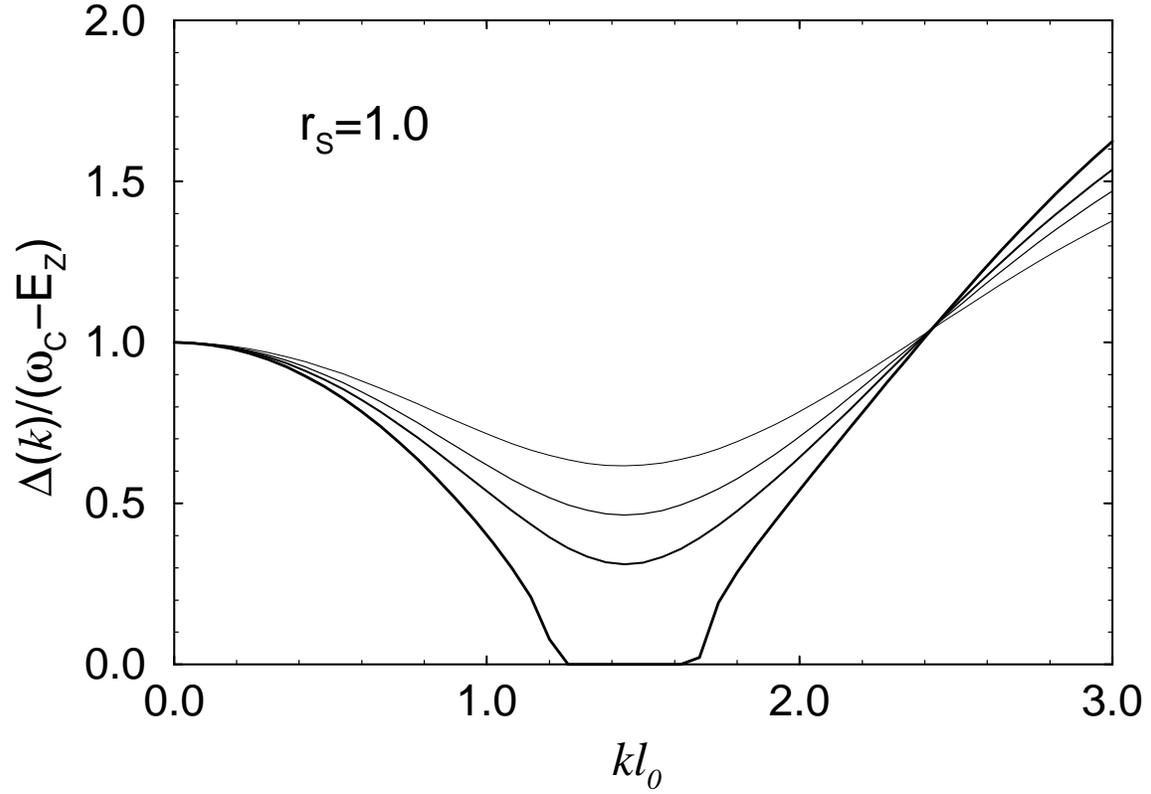,width=7.0in,angle=-90}}
\caption{ Dispersion curves of the lowest spin density excitation
in the unpolarized IQHE state at $\nu =2$ for $r_S =1.0$ and
various $E_{Z}$'s. In ascending order of line thickness
the values of  $E_Z/\hbar\omega$ are 0.50, 0.60, 0.65 and 0.70.
\label{unsde}}
\end{figure}

The dispersion curves of the above pure modes are plotted in Fig.\ref{pureun}.
We can use Kohn's theorem to check if our approximation is reasonable.
Fig.\ref{pureun} shows that the energy of charge density excitation
approach $\omega_C$ as $k \rightarrow 0$.
Since the charge density excitation does not show any instability, we
will concentrate only on the spin density excitation to
obtain the phase boundary which is the critical Zeeman splitting energy
needed for the stable excitation.  
\mbox{Fig.\ref{unsde}} shows the dispersion curves of the lowest
spin density excitation for $r_S = 1.0$ and various values of $E_Z$.

\section{The phase diagram}
\label{sec-phase}

The phase diagram of the state at $\nu=2$ can be obtained at two levels of
sophistication.

\subsection{Phase diagram in Hartree Fock approximation}

The simplest approximation is that of non-interacting electrons.  In this case,
the phase boundary is given simply by $E_Z=\hbar\omega_C$, 
as shown in \mbox{Fig.\ref{fullphase}}. 
As we shall see, this is sensible only in the limit of $r_S\rightarrow 0$;
interactions modify the phase diagram substantially elsewhere.

In the simplest approximation, interaction can be incorporated by comparing the
energies of the fully polarized and unpolarized ground states in the Hatree-Fock
approximation, that is, by assuming 
that the ground state contains either 0$\uparrow$
and 0$\downarrow$ Landau levels fully occupied or 0$\uparrow$ and 1$\uparrow$.
The contributions of the kinetic energy and the Zeeman
coupling to the ground state energy are straightforward.
The exchange interaction energy in the Hartree-Fock
approximation can be evaluated in terms of the self-energy defined
in the previous section. The self-energy $\Sigma^{n(\sigma)}_{n}$ 
is the exchange interaction
between an electron in the Landau level with index $n$
and all the other electrons with the same spin.
The exchange interaction energy per particle is then the sum of 
the self-energies for all electrons 
divided by two times the number of electrons where the factor of two
prevents a double counting. That is to say,




\begin{equation}
V_{ex} = \frac{1}{2\nu}
\sum^{1/2}_{\sigma= -1/2} \Big( \sum^{n(\sigma)}_{n' = 0} 
\Sigma^{n(\sigma)}_{n'} \Big).
\end{equation}
Therefore the ground state energy per particle is

\begin{equation}
E_g = \frac{1}{\nu}
\sum^{1/2}_{\sigma= -1/2} \sum^{n(\sigma)}_{n' = 0} 
\Big[
(n'+\frac{1}{2})\hbar\omega_C +\sigma E_Z
+\frac{1}{2}\Sigma^{n(\sigma)}_{n'} 
\Big].
\end{equation}
The average energy of the fully polarized ground state is

\begin{eqnarray}
\frac{E_g(2:0)}{\hbar\omega_C} -\frac{1}{2}= 
\frac{1}{2} - \frac{1}{2}\frac{E_Z}{\hbar\omega_C}
+\frac{1}{4} (\tilde{\Sigma}^{1}_{n=0} +
\tilde{\Sigma}^{1}_{n=1}) r_S
= \frac{1}{2} - \frac{1}{2}\frac{E_Z}{\hbar\omega_C}
-\frac{11}{16} \sqrt{\frac{\pi}{2}} r_S.
\end{eqnarray}
Similarly the average energy of the unpolarized ground state is

\begin{eqnarray}
\frac{E_g(1:1)}{\hbar\omega_C} -\frac{1}{2}= 
\frac{1}{2} \tilde{\Sigma}^{0}_{n=0} r_S
= -\frac{1}{2} \sqrt{\frac{\pi}{2}} r_S.
\end{eqnarray}
The phase boundary is given by the solution of the following equation:
\begin{eqnarray}
\frac{E_g(2:0)-E_g(1:1)}{\hbar\omega_C} = 0.
\end{eqnarray}
Therefore the critical $E_Z/\hbar\omega_C$ as
a function of $r_S$ is

\begin{eqnarray}
\frac{E_Z}{\hbar\omega_C} = 1 - \frac{3}{8}\sqrt{\frac{\pi}{2}} r_S  
\end{eqnarray}
shown in \mbox{Fig.\ref{fullphase}}.

\subsection{Phase diagram from collective mode instability}

The phase diagram obtained by a comparison of the energies of the Hartree Fock 
states 
is not fully reliable for two reasons.  First, it neglects Landau level
mixing, which is crucial for the issue of interest here.  Secondly,
it does not allow for the possibility of other states in the phase diagram.  
Therefore, it is
more appropriate to look for instabilities of the two 
states by asking when one of the collective modes becomes soft.
Indeed, a first order phase transition may (and likely will) occur 
even before the collective mode energy approaches zero, but we believe that 
the phase diagram obtained by considering instabilities ought to be 
reliable qualitatively and even semi-quantitatively.

\begin{figure}
\centerline{\psfig{figure=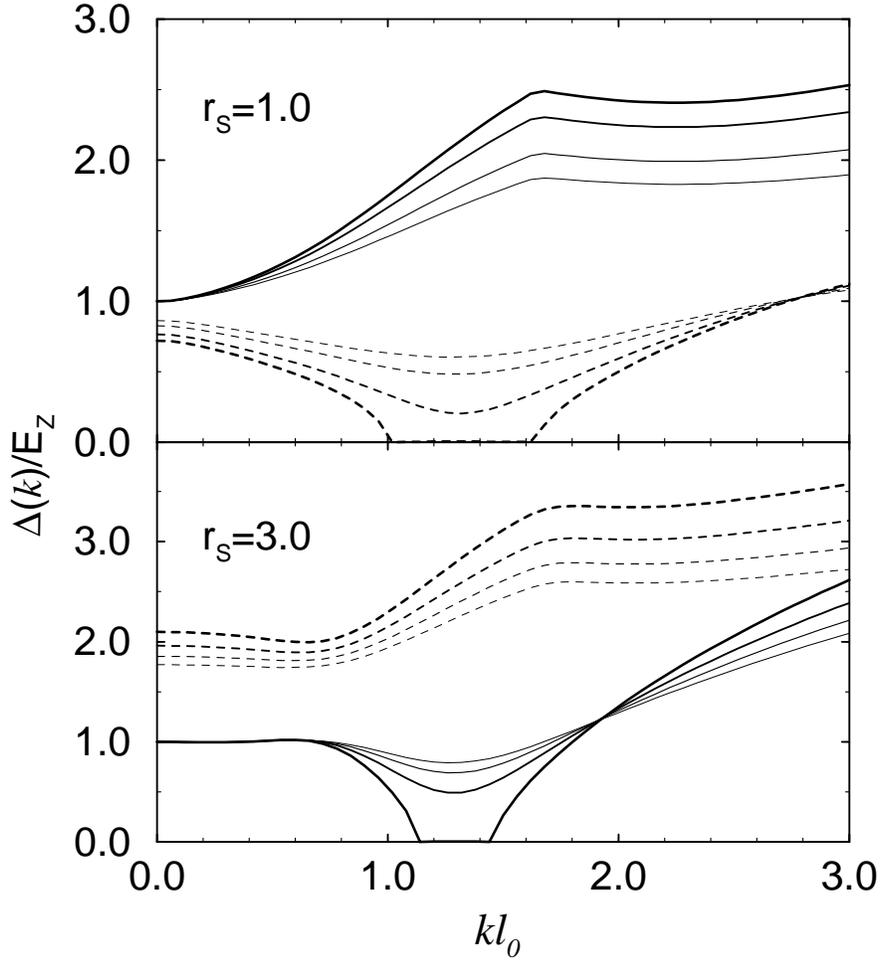,width=5.0in,angle=0}}
\caption{ Dispersion curves for the spin wave excitation and the 
$m = -1$ spin-density  mode
in the fully polarized IQHE state at $\nu =2$ for $r_{S} = 1.0$ and $3.0$.
The solid line corresponds to the spin wave excitation while the dashed line
to the $m = -1$ mode.  The various dispersion curves  
are plotted for different ratios of $E_Z/\hbar\omega$; in ascending order of
line thickness, this ratio is given by 1.2, 1.0, 0.8
and 0.7 in the top graph and by 1.8, 1.6, 1.4 and 1.2 in the bottom graph. 
The kinks in the dispersion appear at the anticrossings between different
modes.  (Only the lowest two collective modes are shown here.)
While the $m=-1$ mode has lower energy at $r_S=1$, the spin wave mode is the
lowest energy mode for $r_S=3$.
\label{polsde}}
\end{figure}

As noted previously, there is no instability in the charge density wave
collective mode in the parameter range considered here. 
We have determined the onset of the spin density collective mode
instability  both in the fully polarized and the unpolarized states by varying 
$r_S$ and $E_Z$, as shown in \mbox{Fig.\ref{polsde}} and 
\mbox{Fig.\ref{unsde}}.  
\mbox{Fig.\ref{fullphase}} shows the phase diagram thus obtained.
The following features are noteworthy.

(i) The nature of instability is different depending on whether $r_S$  
is small or large.
At small $r_S$, where the interactions are negligible and the physics is
dictated by the Zeeman energy, the lowest energy
spin-density excitation is clearly the $m=-1$ mode.
This continues to be the case for  
$r_S\lesssim 2$; here the $m=-1$
spin-density mode is responsible for the instability of the (2:0) state,
as shown in \mbox{Fig.\ref{polsde}}.
However, for $r_S \gtrsim 2$, interactions are sufficiently strong that the
spin-wave mode becomes the lowest energy mode and causes
the instability.  One way to understand why the $m=0$ spin wave 
mode has lower energy at large $r_S$ than the $m=-1$ mode 
is because whereas the former has an energy equal to $E_Z$ in the
long wave length limit no matter what $r_S$,  
as guaranteed by the Goldstone theorem, 
the energy of the latter is determined by the interactions.

(ii) At small $r_S$, the interactions make the fully polarized (2:0) state 
more stable as compared to the non-interacting problem, 
as evidenced by the fact that the transition out of it takes place
at a Zeeman energy {\em smaller} than $\hbar\omega_C$.
This is precisely as expected from the exchange physics, as also 
captured by the Hartree-Fock phase diagram.

(iii)  The instability in the spin-wave mode occurs through the development 
of a roton minimum, the energy of which vanishes at certain Zeeman energy.  
The roton minimum in turn is caused by the Landau level mixing, underscoring
the important role of Landau level mixing at large $r_S$.
Without Landau level mixing the spin-wave mode does not show
any instability as shown in \mbox{Fig.\ref{puresde}}.

\begin{figure}
\centerline{\psfig{figure=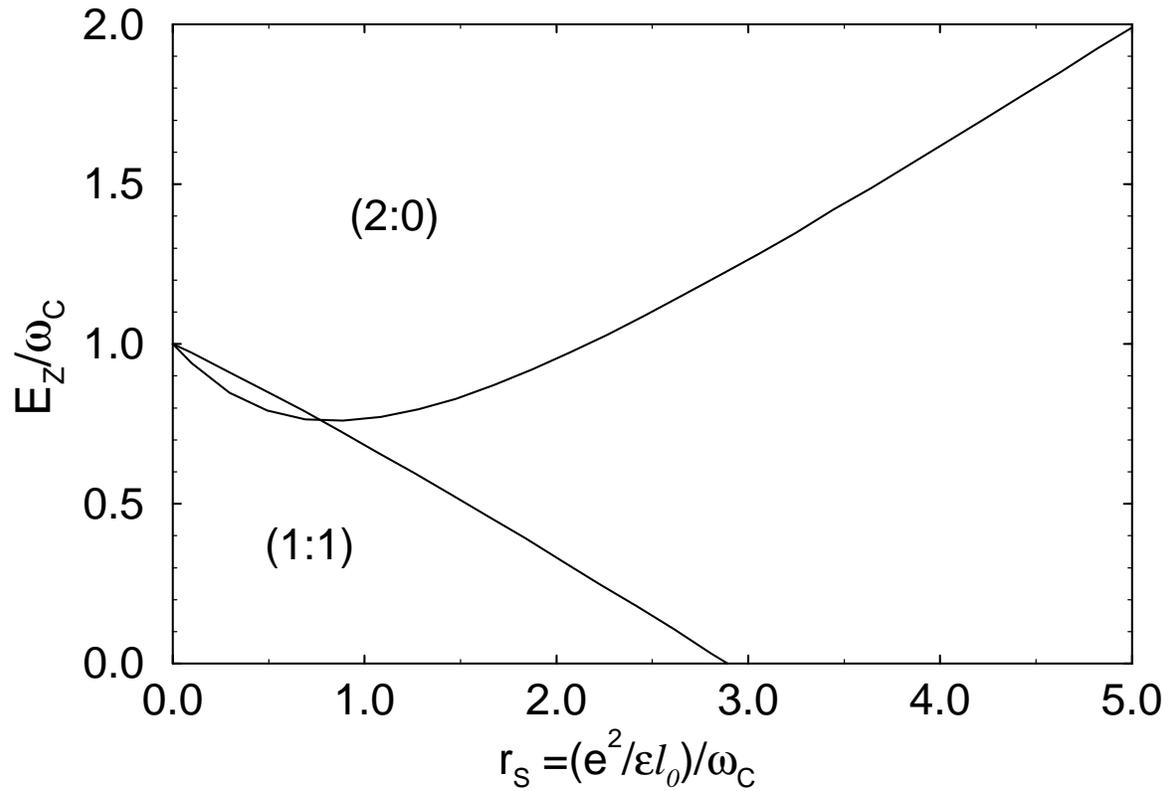,width=7.0in,angle=-90}}
\caption{ Phase diagram of the $\nu =2$ state as
a function of $E_Z/\omega_C$ and $r_S$.
The phase boundaries are computed from the roton instability of
the spin wave excitation of the fully polarized and unpolarized state. 
The region where the fully polarized state
is stable is denoted by (2:0) while the region for the unpolarized state
is denoted by (1:1).
\label{phase}}
\end{figure}

(iv)  For the unpolarized state, the lowest spin density excitation is the 
$m=1$ mode at all $r_S$, the long wavelength limit of the excitation energy of
which is fixed to be $\hbar\omega_C - E_Z$ independent of $r_S$.  
Here also, the instability occurs through a roton, which becomes deeper as the
$r_S$ is increased (and interactions become stronger).  Since a spin flip is
favored by exchange, the energy of the $m=1$ mode decreases with increasing
$r_S$, consistent with the  feature that the critical $E_Z$ in this case is
monotonically decreasing as a function of $r_S$,
as is shown in \mbox{Fig.\ref{phase}}.

(v) At small $r_S$ there is a small region in \mbox{Fig.\ref{phase}}
where the two phases coexist.  This clearly is an artifact of various
approximations involved in our calculation, and
implies that the actual locations of the phase boundaries cannot be taken too
seriously.  As mentioned earlier, a first order transition is likely to occur
{\em before} the roton energy vanishes.

(vi) At large $r_S$ and small $E_Z$, the ground state is derived neither from
(2:0) nor from (1:1).  Since we find a {\em finite} wave vector instability   
in the spin density wave excitation, it is natural to expect that the state
here has a spin-density wave state.  Further work will be required to establish
the nature of this state in more detail.

\section{Conclusion}
\label{sec-conclusion}

\begin{figure}
\centerline{\psfig{figure=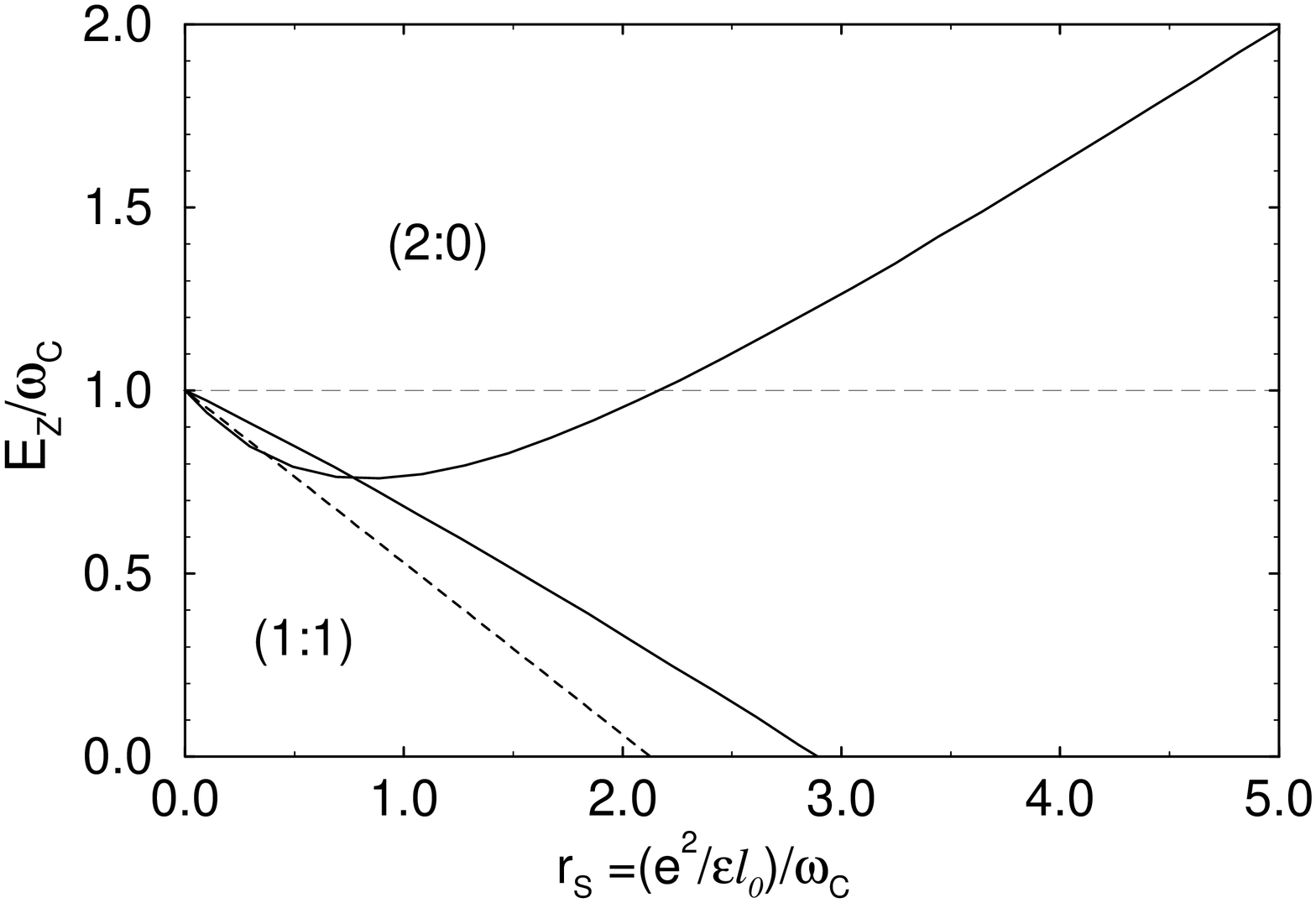,width=7.0in,angle=-90}}
\caption{ Phase diagram of the $\nu =2$ state as
a function of $E_Z/\omega_C$ and $r_S$. The solid lines indicate 
the phase boundaries which are computed from the roton instability of
the fully polarized and unpolarized states. The dashed line indicates
the phase boundary obtained from the comparison between the two 
ground state energies in the Hartree-Fock approximation. 
The long-dashed
line is the phase boundary of non-interacting electron system.
The region where the fully polarized state
is stable is denoted by (2:0) while the region for the unpolarized state
is denoted by (1:1).  
\label{fullphase}}
\end{figure}

The principal outcome of our calculations is the phase diagram in 
\mbox{Fig.\ref{fullphase}}
which shows the regions where the fully polarized and the unpolarized Hatree 
Fock states (2:0) and (1:1) are valid.
We believe that it should be possible to investigate 
the roton minimum in the spin-wave excitation of the fully polarized state
as well as its instability in inelastic light scattering experiments.

Another situation where similar physics may apply is in the case of composite
fermions \cite{Jain} at effective filling factor $\nu^*=2$, which corresponds
to the electron filling factor $\nu=2/5$.  It is easier in this case to see a
transition between  the fully polarized and the unpolarized states because the
effective cyclotron energy for composite fermions 
is substantially small compared to the cyclotron energy of electrons, which
makes it possible to obtain $E_Z$ comparable to or larger than the effective
cyclotron energy in tilted field experiments.  The critical Zeeman energy at
the transition was calculated by Park and Jain by comparing the ground state
energies \cite{Park}, in 
reasonable agreement with the experiments of  Kukushkin, von Klitzing, and
Eberl \cite{Kukushkin}.  Park and Jain also estimated a mass for composite
fermions, the ``polarization mass" by equating the critical Zeeman energy to 
the effective cyclotron energy.  There is one subtlety though.  
In the case of composite fermions both the effective cyclotron energy and the
effective interactions derive from the same underlying energy, namely the
Coulomb interaction between the electrons, and therefore neither the 
interactions between composite fermions nor a mixing  between
{\em composite fermion} Landau levels can, in principle, be neglected 
in any realistic limit.  These would provide a correction to the mass obtained
in \mbox{Ref.\cite{Park}}.  
However, we note that the mass was reliable to no more than 
20-30\%, and the corrections may be negligible compared to that.

Interestingly, there is experimental evidence \cite{Du,Kukushkin}
that the transition between the fully polarized and the unpolarized composite 
fermion states does not occur directly but through an intermediate state
with a partial spin polarization.  Murthy \cite{Murthy} has proposed 
that this  state is a Hofstadter lattice of composite 
fermions, and has half the maximum possible polarization.
It would be interesting to see if similar physics obtains for $\nu=2$ as well.
In particular, the phase diagram of 
\mbox{Fig.\ref{fullphase}} 
predicts that for $r_S\gtrsim 1$,
the transition from the fully polarized state to the unpolarized state as a
function of the Zeeman energy is not direct but through another, 
not yet fully identified state.   
(We suspect that this may be true at any arbitrary $r_S$, although not 
captured by our calculated phase diagram.)

This work was supported in part by the
National Science Foundation under grant no. DMR-9986806.
We thank G. Murthy for discussions.

\pagebreak

\pagebreak

\pagebreak

\pagebreak

\pagebreak

\pagebreak

\pagebreak

\pagebreak

\pagebreak

\pagebreak


\begin{thebibliography}{99}

\bibitem{Ceperley} B. Tanatar and D.M. Ceperley, Phys. Rev. B {\bf 39},
5005 (1989).

\bibitem{Pan} W.Pan, J.S. Xia, V. Shvarts, E.D. Adams, H.L. Stormer,
D.C. Tsui, L.N. Pfeiffer, K.W. Baldwin, and K.W. West, cond-mat/9907356.

\bibitem{Pinczuk} M.A. Eriksson , A. Pinczuk , B.S. Dennis , S.H. Simon , 
L.N. Pfeiffer and K.W. West , Phys. Rev. Lett. {\bf 82} , 2163 (1999).


\bibitem{Binding} I.V. Lerner and Yu.E. Lozovik, Zh. Eksp. Teor. Fiz.
{\bf 78}, (1978); Sov. Phys. JETP 
{\bf 51} , 588 (1980).


\bibitem{RPA1} K.W. Chiu and J.J. Quinn, Phys. Rev. B {\bf 9}, 4724 (1974).

\bibitem{RPA2} H. Fukuyama, Y. Kuramoto, and P.M. Platzmann, 
Surf. Sci. {\bf 73}, 491 (1978); Phys. Rev. B {\bf 19}, 4980 (1979)


\bibitem{Halperin} C. Kallin and B.I. Halperin , Phys. Rev. B {\bf 30} , 5655
(1984).

\bibitem{MacDonald1} A.H. MacDonald, cond-mat/9410047.

\bibitem{MacDonald2} A.H. MacDonald, J. Phys. C : Solid State Phys.
{\bf 18} , 1003 (1985).

\bibitem{Kallin} C. Kallin, in \emph{Interfaces, Quantum Wells, and
Superlattices}, edited by C.R. Leavens and R. Taylor (Plenum, New York,
1988), p. 163.

\bibitem{thickness} F.F. Fang and W.E. Howard, Phys. Rev. Lett. {\bf 16},
797 (1966); F. Stern and W.E. Howard, Phys. Rev. {\bf 163}, 816 (1967)

\bibitem{Mahan} G.D. Mahan, \emph{Many-Particle Physics}
(Plenum, New York, 1981), p. 133.


\bibitem{NumRecipes} W.H. Press, S.A. Teukolsky, W.T. Vetterling, 
and B.P. Flannery,
\emph{Numerical Recipes in C}, 2nd Edition
(Cambridge University Press, 1992), p. 482.

\bibitem{Kohn} W. Kohn, Phys. Rev. {\bf 123}, 1242 (1961).

\bibitem{Jain} J.K. Jain, Phys. Rev. Lett. {\bf 63}, 199 (1998);
Phys. Rev. B {\bf 41}, 7653 (1990); Science {\bf 266}, 1199 (1994).

\bibitem{Park} K. Park and J.K. Jain, Phys. Rev. Lett. {\bf 80},
4237 (1998).

\bibitem{Du} R.R. Du, A.S. Yeh, H.L. Stormer, D.C. Tsui, L.N. Pfeiffer,
and K.W. West, Phys. Rev. Lett. {\bf 75}, 3926 (1995).

\bibitem{Kukushkin} I.V. Kukushkin, K. von Klitzing, and K. Eberl,
Phys. Rev. Lett. {\bf 82}, 3655 (1999).

\bibitem{Murthy} Ganpathy Murthy, cond-mat 9906110.




\end{thebibliography}
\end{document}